\documentclass{article}
\usepackage{hyperref}
\usepackage{amsmath}
\usepackage{graphicx}
\usepackage{tikz}
\usepackage{float}
\usetikzlibrary{positioning, calc, decorations.pathreplacing}
\usepackage[utf8]{inputenc}
\usepackage[top=1in,bottom =1in, left=1in,right=1in]{geometry}
\usepackage{amsmath,amssymb,mathtools,bm,isomath,amsthm}
\usepackage{graphicx}
\usepackage{epstopdf}
\usepackage{color}
\usepackage{soul}
\usepackage{enumerate}
\usepackage{nicefrac}
\usepackage{enumitem}
\usepackage{mathrsfs}
\usepackage{scalerel,stackengine}
\usepackage{afterpage}
\usepackage{graphicx}
\usepackage{url}
\usepackage[affil-it]{authblk} 
\usepackage[toc,page]{appendix}
\usepackage{hyperref}
\usepackage{amsmath}
\usepackage{tikz}
\usepackage{float}
\usepackage{optidef}
\usepackage{algpseudocode}
\usetikzlibrary{positioning}
\usetikzlibrary{positioning, calc, decorations.pathreplacing}
\usetikzlibrary{matrix, positioning}
\usetikzlibrary{fit,positioning}
\usepackage[export]{adjustbox}
\usepackage{pgfplots}
\usepackage{graphicx}
\usepackage{epstopdf}  
\usepackage{flafter,placeins}
\usepackage{subcaption} % figures & sous figures / figures & subfigures
\usepackage{listofitems}
\usepackage{tikz}
\usepackage{siunitx} % unites SI / SI units
\usepackage{amssymb} % autres symboles mathematiques / other mathematical symbols
\usepackage[bottom]{footmisc} % pour avoir les notes de bas de page en dessous des figures... / to have the footnotes below the figures

\usepackage{amsmath,color,soulutf8,longtable,colortbl,setspace,xspace,url,pdflscape,cite}

\usepackage[nolist]{acronym}

\usepackage{amsmath,color,soul,longtable,colortbl,setspace,xspace,url,pdflscape,cite}

\usepackage[nolist]{acronym}
\usepackage{fancyhdr}
\fancypagestyle{pagenumber}{\fancyhf{}\fancyhead[R]{\thepage}}

\makeatletter
\let\ps@plain=\ps@pagenumber
\makeatother

\usepackage{hyperref}
\makeatletter
\providecommand*{\toclevel@compteur}{0}
\makeatother
\usepackage{caption}
\usepackage{subcaption}
\usepackage{siunitx}
\usepackage{algorithm}  % For algorithm environment
\usepackage{amssymb} 
\usepackage[bottom]{footmisc} 
\newcommand{\x}{\boldsymbol{x}}

\pgfplotsset{compat=1.18} 
\begin{document}
	\setlength{\baselineskip}{15pt}
	
\title{Physics-Informed Neural Networks for the Numerical Modeling of Steady-State and Transient Electromagnetic Problems with Discontinuous Media}
\author[1]{Michel Nohra}
	\author[1]{Steven Dufour}
	\affil[1]{D\'epartement de math\'ematiques et de g\'enie industriel, Polytechnique Montr\'eal, Montr\'eal, Qu\'ebec, Canada, H3T 1J4}
	
	\date{\today}
	
	\maketitle

\begin{abstract}

Physics-informed neural networks (PINNs) have emerged as a promising numerical method based on deep learning for modeling boundary value problems, showcasing promising results in various fields. In this work, we use PINNs to discretize three-dimensional electromagnetic, parametric problems, with material discontinuities, covering both static and transient regimes. By replacing the discontinuous material properties with a continuous approximation, we eliminate the need to directly enforce interface conditions. Using the Neural Tangent Kernel (NTK) analysis, we show that using the first-order formulation of Maxwell's equations is more suitable for interface problems. We introduce a PINN-based decomposition on overlapping domains to enhance the convergence rate of the PINN.

  \end{abstract}

\section{Introduction}
Electromagnetism is a branch of physics that studies the behavior of electric and magnetic fields. It plays an important role in our daily lives, from electrical appliances to electricity generation.
Numerical methods are valuable for modeling and simulating these physical problems~\cite{num_ele1,num_ele2,num_ele3,num_ele4}, allowing the conduct of cheap experiments for driving innovations. 
Recently, neural networks have gained popularity due to advancements in computer hardware, and their ability to model complex systems \cite{ann}. They have been used in a wide range of sectors, from computer vision to speech recognition \cite{nn_1,nn_2,nn_3}. More recently, they have been used for scientific computing in numerous applications, such as for building turbulence models from data \cite{turb1, turb2,turb3,turb4}, and for studying heat transfer problems~\cite{HT1,ht2}. Among the methods that use deep learning for scientific computing, Physics-Informed Neural Networks (PINNs)\cite{raissi} are becoming popular among computational scientists. PINNs, first introduced by Raissi et al.\cite{raissi}, incorporate the Partial Differential Equations (PDEs), and all the available information on their solution, for training the neural network to approximate the solutions. The boundary-value problem is then transformed into an optimization problem.
Since their introduction, PINNs have been used for various forward, inverse, and parametric problems. To list a few applications,  researchers have used PINNs with heat transfer problems to estimate velocity and temperature fields, and in cases where parts of the boundary conditions are missing, but values of the dependent variables at specific interior points are known \cite{HT1}. 
In fluid dynamics, PINNs have been used with the velocity-pressure and the vorticity-velocity formulations to reproduce benchmark problems\cite{NSFnet}. PINNs were also used to approximate the level-set function in multifluid flows in \cite{nohra1}.
In electromagnetism, PINNs have been used to address both forward and inverse problems.
As for electromagnetism, PINNs have been used to address both forward and inverse problems.
A PINN is used by Kovacs et al. \cite{max2} for forward and inverse 2D magnetostatic problems. The forward problem aims at approximating the magnetic vector potential, given a magnetization distribution. The inverse problem aims at approximating the magnetization distribution, given a magnetic vector potential. The authors' methodology uses two neural networks. The first network takes the spatial coordinates as inputs, and it is trained to approximate the vector potential for a given magnetization distribution. The second network takes the spatial coordinates as inputs, it gives the magnetization as an output, and it is trained to find the magnetization that, when used with the first neural network, produces the desired vector potential.
Lim et al. \cite{max_net} approximate the electric field distribution for a given electromagnetic lens shape, using a PINN trained with the 2D $E$-formulation in the frequency domain. The inverse problem is also addressed, where they look for the lens shape that produces a given electric field distribution. The authors use a decoder network to parameterize the lens shape, where the input is a latent vector (a vector of relatively small dimension), and the output is the lens shape. The values of the latent vector are then optimized to produce the desired electric field.
A neural network was used by Fang et al. \cite{meta_des} to approximate the $z$-component of an electric field using the 2D $E$-formulation in the frequency domain, for given permeability and permittivity. The inverse problem was also addressed, where a second neural network is used to approximate the permeability and permittivity that produce a given electric field when fed to the first network.

Limited work has been done on the use of PINNs for interface problems. Tseng et al.~\cite{tseng2023cusp} addressed the interface problem for elliptic PDEs, by including the distance to the interface in the input of the neural network. Since the distance function has a discontinuous derivative at the interface, the network's output should also have a discontinuous derivative. The authors included interface conditions, such as the jump of the first-order derivatives, in the loss function. 
Wu et al. and Zhu et al.~\cite{wu2022inn,zhu2023physics} partitioned the domain into subdomains along the interface, where different neural networks were used to approximate the solution in each subdomain. Interface conditions were included in the loss function.

In this study, we use a PINN for modeling electromagnetic three-dimensional transient problems with discontinuous media. 
The PINN is trained with the first-order Maxwell's equations in the low-frequency regime, with strongly imposed boundary conditions. A numerical study is conducted to identify the optimal choice of functions to be used for the boundary conditions. Additionally, a novel neural network architecture is implemented to enhance the convergence of the PINN.
The verification is performed on 3D, parametric, static and transient problems.

We first give a brief introduction to Maxwell's equations. We then describe the PINN method. We explain the choice of the formulation of Maxwell's equations using the Neural Tangent Kernel (NTK). We then propose a neural network's architecture that is suitable for the interface problem, and we finally verify and validate the proposed methodology on multiple $3$D parametric and transient problems.

\section{Maxwell's equations}

Maxwell's equations are a set of four partial differential equations that describe the dynamics of electromagnetic and electric fields,
and how they interact with each other, and with matter. They form the basis for many technologies we use everyday, such as computer chips and antennas.
For linear materials, the relations between the magnetic field $\boldsymbol{H}$, the magnetic flux $\boldsymbol{B}$, the electric field $\boldsymbol{E}$, and the electric displacement $\boldsymbol{D}$, are given by:
\begin{subequations}
\begin{align}
    \partial_t \mu\boldsymbol{H}+ \nabla \times\boldsymbol{E}= 0; \label{eqn:full_max_21}\\
    \partial_t \epsilon\boldsymbol{E}- \nabla \times\boldsymbol{H}+\boldsymbol{J}= 0; \label{eqn:full_max_22}\\
    \nabla \cdot \epsilon\boldsymbol{E}= \rho_c; \label{eqn:full_max_23}\\
    \nabla \cdot \mu\boldsymbol{H}= 0, \label{eqn:full_max_24}
\end{align}
\label{eqn:full_max_2}
\end{subequations}
\vspace{-\baselineskip}

with $\boldsymbol{B}=\mu \boldsymbol{H}$ and $\boldsymbol{D}=\epsilon \boldsymbol{E}$, where the permeability $\mu$ and permittivity $\epsilon$ are constants, where $\rho_c$ is the charge density, and $\boldsymbol{J}$ is the current density given by $\boldsymbol{J}= \sigma\boldsymbol{E}+ \boldsymbol{u} \times \mu \boldsymbol{H},$ where $\sigma $ is the conductivity, and $\boldsymbol{u}$ is the velocity of the electric charge. In this work, we will consider this velocity to be zero.

\subsection{Maxwell's equations formulations}
\label{sec:VFM}
Various formulations exist for Maxwell's equations, based on different assumptions, and for various applications. We start with the steady-state assumption, where we consider that the dependent variables do not vary in time, i.e. $\frac{\partial \cdot}{\partial t} =0$. System~\eqref{eqn:full_max_2} then becomes:
\begin{subequations}
\begin{align}
    \nabla \times\boldsymbol{E}=0 ; \label{eqn:steady_max1}\\
    \nabla \times\boldsymbol{H}= \sigma\boldsymbol{E}; \label{eqn:steady_max2}\\
    \nabla \cdot \epsilon\boldsymbol{E}= \rho_c; \label{eqn:steady_max3}\\
    \nabla \cdot \mu\boldsymbol{H}= 0.\label{eqn:steady_max4}
\end{align}
\label{eqn:steady_max}
\end{subequations}

From equation~\eqref{eqn:steady_max1}, using the Helmholtz decomposition, we have that $\boldsymbol{E}=\nabla V$, where $V$ is a scalar field, usually called the electric potential. For conductive media, where $\boldsymbol{J} \neq 0$, we obtain
$$\nabla \cdot\boldsymbol{J}= \nabla \cdot \sigma\boldsymbol{E}  =\nabla \cdot \sigma \nabla V =0 , $$ which is known as the electric potential field formulation for conductive media. 
As for the case of perfectly insulating materials, where $\boldsymbol{J} = 0$, we obtain from equation~\eqref{eqn:steady_max3}, 
$$\nabla \cdot \epsilon \nabla V =\rho_c. $$
For the case of perfectly insulated materials, and when the field of interest is $\boldsymbol{H}$, equation \eqref{eqn:steady_max2} gives $$\nabla \times\boldsymbol{H}=0.$$ Hence, using the Helmholtz decomposition, we obtain that $\boldsymbol{H}=\nabla A$, where $A$ is a scalar field called the magnetic scalar potential. The property that $\mu\boldsymbol{H}$ 
 is divergence-free gives us an equation for $A$,
$$\nabla \cdot \mu \nabla A =0 .$$ 
In conductive media, and when $\boldsymbol{H}$ is the field of interest, equation~\eqref{eqn:steady_max3} implies that \\$\mu\boldsymbol{H}= \nabla \times \boldsymbol{A} ,$ and we obtain, from equation~\eqref{eqn:steady_max2},
$$\nabla \times \frac{1}{\mu} \nabla \times \boldsymbol{A} = \boldsymbol{J}. $$

As for the low-frequency regime assumption, where $\partial_t \epsilon\boldsymbol{E}<<\boldsymbol{J}$, the following quasi-static equations are obtained from system~\eqref{eqn:full_max_2}:
\begin{subequations}
\begin{align}
  \partial_t \mu\boldsymbol{H}+ \nabla \times\boldsymbol{E}= 0; \label{eqn:lowf_max_21}\\
   - \nabla \times\boldsymbol{H}+ \sigma\boldsymbol{E} =0; \label{eqn:lowf_max_22}\\
  \nabla \cdot \mu\boldsymbol{H}=0; \label{eqn:lowf_max_23}\\
  \nabla \cdot \epsilon\boldsymbol{E}= \rho_c. \label{eqn:lowf_max_24}
\end{align}
\label{eqn:lowf_max_2}
\end{subequations}

A common formulation for the quasi-static equations is obtained by replacing $\boldsymbol{E}$ from equation~\eqref{eqn:lowf_max_22} in equation~\eqref{eqn:lowf_max_21}, which gives
\begin{equation}
\begin{split}
  \partial_t \mu\boldsymbol{H}+ \nabla \times \frac{1}{\sigma} \nabla \times\boldsymbol{H} = \boldsymbol{F}_{\rm ext}; \\
  \nabla \cdot \mu\boldsymbol{H}=0, 
\end{split}
\label{eqn:H-form_T2}
\end{equation} 
 and is usually referred to as the $H$-formulation. 
A similar approach for $\boldsymbol{E}$ yields
$$ \mu \partial_t (\sigma \boldsymbol{E}) +  \nabla \times  \nabla \times\boldsymbol{E}=0.$$

\subsection{Interface conditions}
In the presence of a material interface,  $\mu$, $\epsilon$, and $\sigma$ are discontinuous. This leads to the following interface conditions:
\begin{itemize}
    \item From the divergence-free property $\nabla \cdot \mu\boldsymbol{H}= 0,$ we obtain that $$\mu_1 \boldsymbol{H}_{1n} = \mu_2 \boldsymbol{H}_{2n},$$ where $\boldsymbol{H}_{1n}$ denotes the normal component of $\boldsymbol{H}$ on the first side of the interface,  and $\boldsymbol{H}_{2n}$ is the normal component on the second side of the interface;
    \item By applying the divergence operator on equation~\eqref{eqn:lowf_max_22}, we obtain $$\nabla \cdot(- \nabla \times\boldsymbol{H}+ \sigma \boldsymbol{E})  = \nabla \cdot \sigma\boldsymbol{E}= 0 ,$$ leading to the interface condition 
    $$\sigma_1 \boldsymbol{E}_{1n} = \sigma_2 \boldsymbol{E}_{2n};$$ 
    \item  From equation~\eqref{eqn:lowf_max_22}, we obtain that $\boldsymbol{H}_{1t} - \boldsymbol{H}_{2t} = \boldsymbol{K}_f$, where $\boldsymbol{K}_f$ is the surface current density on the interface, and $\boldsymbol{H}_{1t}$ and $\boldsymbol{H}_{2t}$ are the tangential components of $\boldsymbol{H}$ on both sides of the interface;
    \item From equation~\eqref{eqn:lowf_max_21}, we obtain $\boldsymbol{E}_{1t} = \boldsymbol{E}_{2t}.$
\end{itemize}
      
\section{Numerical methodology}

\subsection{Physics-informed neural networks}

\begin{figure}
    \centering
    \includegraphics{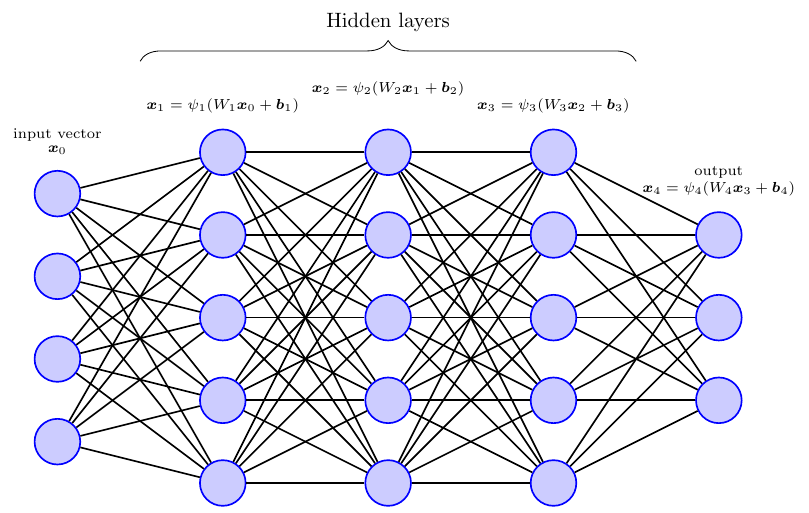}
    \caption{Feedforward neural network}
    \label{fig:FFNN_T2}
\end{figure}

To better understand neural networks, let us look at a simple example of a one-layer neural network. If we consider the input vector \(\x\) and the output vector \(\x_1\) to have dimensions \(n\) and \(m\) respectively, then the output is given by
\[\x_1 = \psi (W\x + \boldsymbol{b}),\]
where \(W\) is an \(m \times n\) weight matrix and \(\boldsymbol{b}\) is a bias vector of dimension \(m\). The activation function \(\psi(\boldsymbol{z})\) is user-defined,  which could be, for example, a cosine or a hyperbolic tangent function, and is applied to each component of a vector \(\boldsymbol{z}\).
By stacking layers, we create what is known as a feedforward neural network, which involves
passing the output of one layer to the input of the next layer, as illustrated in figure~\ref{fig:FFNN_T2}. In this setup, the final output of the network is described by
\[ N(\x) = \psi_\ell \big( W_\ell  \psi_{\ell-1}( W_{\ell-1} \psi_{\ell-2}(\cdots (\psi_1(W_1 \x + \boldsymbol{b}_1  )  )   \cdots) +\boldsymbol{b}_{\ell-1}  )       +\boldsymbol{b}_\ell     \big).  \]

\tikzstyle{mynode}=[thick,draw=blue,fill=blue!20,circle,minimum size=22]

One can build a more complex neural network by changing the connections (the architecture), between the layers.
After the neural network is set up with its specific architecture and activation functions, the next step is to calibrate its parameters $\boldsymbol{\zeta},$ a vector that contains all the
weights and biases, so that it produces the desired output. This calibration process, commonly referred to as training, is treated as an optimization problem, where we minimize a user-defined loss function \(L\),
\begin{equation*}
    \min_{{\boldsymbol{\zeta}}}\ L(N_{\boldsymbol{\zeta}}(\x), D(\x)),
\end{equation*}
where \(N_{\boldsymbol{\zeta}}(\x)\) denotes the output of the neural network for a given $\boldsymbol{\zeta}$, \(D(\x)\) represents the desired output, and \(L\) is the loss function.
Examples of loss functions are the Mean Squared Error (MSE), often used for regression problems, and the Cross-Entropy, for classification problems.

While various optimization methods could address this minimization problem, in the context of this work, we will only consider gradient-based optimization methods.
Gradient-based methods leverage the property that the gradient of a function points in the direction of the function's steepest ascent. Therefore, to minimize a function, one would move in the direction opposite to the gradient, and the general evolution of the parameters ${\boldsymbol{\zeta}}$ during training is expressed as
$${\boldsymbol{\zeta}}_s ={\boldsymbol{\zeta}}_{s-1}  -d(\alpha) \nabla L,$$
where $d$ is the step size, and ${\alpha}$ is a user-defined parameter, usually called the learning rate.

The two gradient-based methods that we will use are the Adaptive Moment Estimation (Adam) method~\cite{adam} and the Limited-memory Broyden-Fletcher-Goldfarb-Shanno (L-BFGS) method \cite{bollapragada2018progressive}.
 The gradient of $N_{\boldsymbol{\zeta}}(\x)$ with respect to ${\boldsymbol{\zeta}}$ and $\x$, which is needed for these minimization methods, is calculated using the back-propagation algorithm with automatic differentiation.

Let us consider a domain $\upOmega$ with the boundary $\mathrm{\Gamma} = \partial \upOmega .$ We want to find an approximation of the solution $u(\x)$ of a boundary-value problem:
\begin{equation*}
    \begin{aligned}
         &R(\x,u ) =0\ , \ \forall \x \in \upOmega ;\\
         &B(\x ,u) =0\ , \ \forall \x \in \mathrm{\Gamma},
    \end{aligned}
\end{equation*}
where $R$ is the residual of a PDE, and $B$ is the boundary condition on $u(\x)$ we want to impose on ${\rm \Gamma}.$
With the original PINN \cite{raissi}, we approximate $u(\x)$ using a feedforward neural network $N_{\boldsymbol{\zeta}}(\x)$, and we tune the parameters ${\boldsymbol{\zeta}}$ by minimizing the loss function
\begin{align*}
L &= \alpha_1 \sum_{i=1}^{{n}_\upOmega} R(\x_\upOmega^i, N_{\boldsymbol{\zeta}}(\x_\upOmega^i) )^2 + \alpha_2 \sum_{j=1}^{{n}_\mathrm{\Gamma}} B (\x_\mathrm{\Gamma}^j, N_{\boldsymbol{\zeta}}(\x_\mathrm{\Gamma}^j) )^2 + \alpha_3 \sum_{k=1}^{{ n}_{\rm D}}( N_{\boldsymbol{\zeta}}(\x_{\rm D}^k) - u(\x_{\rm D}^k) )^2  \\
&= \alpha_1 \|  R (\x_\upOmega, N_{\boldsymbol{\zeta}}(\x_\upOmega) ) \| + \alpha_2 \|B(\x_\mathrm{\Gamma}, N_{\boldsymbol{\zeta}}(\x_\mathrm{\Gamma}) ) \| + \alpha_3 \| N_{\boldsymbol{\zeta}}(\x_{\rm D}) - u(\x_{\rm D})  \|,
\end{align*}
where $\{\x_\upOmega^i\}_{i=1}^{{ n}_\upOmega}$ is a set of ${ n}_\upOmega$ points that belong to $\upOmega$, 
$\{\x_\mathrm{\Gamma}^i\}_{i=1}^{{ n}_\mathrm{\Gamma}}$ is a set of ${ n}_\mathrm{\Gamma}$ points that belong to $\mathrm{\Gamma}$,
$\{\x_{\rm D}^i\}_{i=1}^{{ n}_{\rm D}}$ is a set of 
${n}_{\rm D}$ points where the solution is known (experimental measurements), $\alpha_i$ are weighting parameters that are user-defined, and $\|.\|$ is the discrete L2 norm.

The original PINN exhibits multiple limitations leading to prolonged training times and poor convergence properties. These include the weak imposition of boundary conditions, the high number of constraints in the loss function which complicates the optimization landscape, and the inherent frequency bias of neural networks \cite{NTK_PINN}. 

Multiple variations of PINNs have been proposed, most notably:
\begin{itemize}

\item PINNs with strong imposition of boundary conditions \cite{sukumar2022exact}: With this method, the neural network's output is multiplied by a function that is zero on the boundaries, and a lift function that satisfies the required boundary conditions is added.
The architecture of the network is shown in Figure \ref{fig:net_arch_T2}, where $d(\boldsymbol{x})$ is the distance function to the boundary, and $V(\boldsymbol{x})$ is the lift function. For complex geometries, where $d(\x)$ is difficult to determine, $d(\x)$ is approximated using a separate neural network.

  \begin{figure}
\begin{center}

\tikzstyle{block} = [draw, fill=white, rectangle, 
    minimum height=3em, minimum width=6em]
\tikzstyle{sum} = [draw, fill=white, circle, node distance=1cm]
\tikzstyle{input} = [coordinate]
\tikzstyle{output} = [coordinate]
\tikzstyle{pinstyle} = [pin edge={to-,ultra thick,black}]

\begin{tikzpicture}[auto, node distance=2cm,>=latex]

    \node [block, name=input] (controller) {$\boldsymbol{x}$ };
    
    \node [block, right of=controller,
            node distance=4cm] (system) {neural network $N(\boldsymbol{x})$};
            
    \node [block, above of=system] (boundary) {boundary function $V(\boldsymbol{x})$};
    \node [block, below of=system] (distance) {distance function $d(\boldsymbol{x})$};
    
    \node [block, right of=system,node distance=5cm] (sol) {$f(\boldsymbol{x})=N(\boldsymbol{x})d(\boldsymbol{x}) + V(\boldsymbol{x})$};

    \draw [->,very thick] (controller) -- node[name=u] {} (boundary);
    \draw [->,very thick] (controller) -- node[name=u] {} (system);
    \draw [->,very thick] (controller) -- node[name=u] {} (distance);
    
    \draw [->,very thick] (system) -- node[name=u] {} (sol);
    \draw [->,very thick] (boundary) -- node[name=u] {} (sol);
    \draw [->,very thick] (distance) -- node[name=u] {} (sol);
  
\end{tikzpicture}

    \centering
    
    \caption{The architecture of the neural network with the strong imposition of the initial condition.}
    \label{fig:net_arch_T2}
\end{center}
\end{figure}
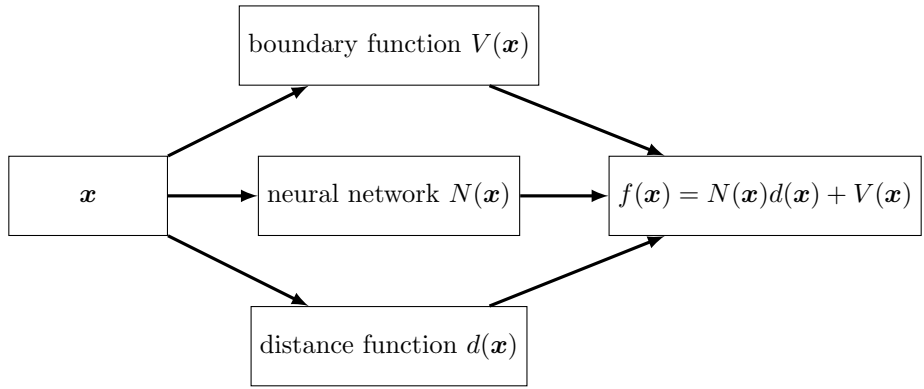

\item PINN with order reduction \cite{FO-PINN}: In some cases, the presence of high-order derivatives slows the convergence during training, and leads to a stiff problem. High-order derivatives can be avoided using the order-reduction technique, where we obtain a set of first-order equations by introducing complementary variables for order reduction. 
For example, let us consider the equation $\partial_x^2 u = f(x)$. We then introduce a second variable $v$, and a complementary equation $v=\partial_x u$ to obtain a set of first-order differential equations of the form:
\begin{align*}
v &= \partial_x u; \\
\partial_x v &= f(x).
\end{align*}
 Order-reduction also reduces the computational cost per training iteration, because the cost of training the extra parameters introduced by the complementary variable is less expensive than calculating the second-order derivative using back-propagation.
 
\item Conservative PINN \cite{cpinn}: Domain decomposition and a separate PINN for each subdomain is used. The continuity of the dependent variables and flux between each subdomain are treated as boundary conditions and are included in the loss function of the associated PINN.
\item Finite-Basis PINN \cite{FBPINN}: Another approach for domain decomposition is to use multiple PINNs defined on overlapping domains~\cite{FBPINN}. A bump function defines each subdomain, and a neural network is assigned to each of them. The final result is obtained by taking the sum of the outputs of the neural networks, each multiplied by its corresponding bump function. Doing so removes the need to impose continuity conditions across subdomains.
\item Gradient-enhanced PINN (G-PINN) \cite{Gpinn}: In this work, the authors added the gradient of the residual to the loss function. This could be seen as a regularization technique, where the neural network is trained to give smoother high-order derivatives. The authors found that the GPINN results in higher accuracy at the expense of computational resources. It was also noted that including the gradient in the loss function could result in lower accuracy if not correctly weighted, and in some cases, prevent the network from converging.

\item Fourier-feature PINN (FF-PINN)~\cite{fourier_net}: The input vector $\x$ is mapped to a higher dimensional vector $\x_{\rm f} = [ \cos{(\x B )}, \sin{(\x B)} ]$,  where $B$ is a random matrix with values between $\omega_{\rm min}$ and $\omega_{\rm max}$. Mapping the input to the desired frequency makes it easier for the network to converge to a solution within that frequency range, effectively surpassing the frequency bias of neural networks.

\item Multi-level PINN\cite{ziad}: In this work, the authors train multiple Fourier-feature PINNs in sequence, where each PINN aims to minimize the residual of the previous sequence. We train the first neural network ${\rm PINN}_1(\omega_1,\x)$ for the large-scale, low-frequency component of the solution using the loss function $L({\rm PINN}_1(\omega_1,\x))$, then we train the second neural network ${\rm PINN}_2(\omega_2,\x))$ for a higher frequency using the loss function $$L({\rm PINN}_1(\omega_1,\x) + k_1 {\rm PINN}_2(\omega_2,\x)),$$where $k_1$ is the approximate norm of the error $k_1= \| Y_{\rm exact} - {\rm PINN}_1(\omega_1,\x) \|$. The authors achieved machine precision for certain problems by using only four levels. 
\end{itemize}

\section{The choice of Maxwell's equations formulation}
As seen in section \ref{sec:VFM}, multiple formulations exist for Maxwell's equations. In this section, we investigate which formulation, the first-order system of equations \eqref{eqn:lowf_max_2} or the second-order system of equations~\eqref{eqn:H-form_T2}, are better suited to train a PINN in the presence of a material interface.
 We will develop the Neural Tangent Kernel (NTK) for the first-order and second-order formulations of a simple 1D test-case, and compare the eigenvalues and eigenvectors obtained with two problem sets. 

\subsection{Neural tangent kernel}
The frequency bias of neural networks, supported by empirical and theoretical evidence~\cite{spec_bias,spec_bias2,NTK}, indicates that a neural network tends to learn the low-frequency components of a function before the high-frequency components during training. In practical terms, the frequency bias allows neural networks to be efficient approximators with inherent filtering capabilities. However, in scenarios where high-frequency components are crucial, the neural network may struggle to converge during training.

Recently, the frequency bias of neural networks has been explained using the Neural Tangent Kernel (NTK) \cite{NTK}. For a regression problem, given a desired output $y(x)$, a set of input points $x_i$, and the neural network output $N({\boldsymbol{\zeta}},x)$, the authors in \cite{NTK} defined the operator $$K_{ij} = \big \langle \partial_{\boldsymbol{\zeta}} N({\boldsymbol{\zeta}},x_i),\partial_{\boldsymbol{\zeta}} N({\boldsymbol{\zeta}},x_j)  \big \rangle,$$ 
where $\big \langle \cdot , \cdot \big \rangle  $ is the inner product.
The authors then showed that, by using a gradient descent method, and as the network's width tends to infinity, the operator $K$ remains constant during training, and the evolution of the neural network $N({\boldsymbol{\zeta}}, x)$ follows the equation $$\partial_t( N({\boldsymbol{\zeta}}, x) - y(x)) = - K (N({\boldsymbol{\zeta}}, x) -y(x) ).  $$ 
Since $K$ is symmetric and semi-definite, we can use the eigendecomposition $K=Q D Q^{T}$,  where $D$ is a diagonal matrix whose diagonal elements are the corresponding eigenvalues, and $Q$ is a square matrix with the eigenvectors in its columns.  This eventually yields
$$Q^T E_{\boldsymbol{\zeta}} = e^{-D t} Q^TE_{\boldsymbol{\zeta}_0}, $$
where $E_{\boldsymbol{\zeta}}=N({\boldsymbol{\zeta}},x) - Y(x).$
This shows that the component of $Q^T Y$ associated with the largest eigenvalue decays faster than those associated with smaller eigenvalues.

 In a similar approach, the authors \cite{NTK_PINN} derived the Neural Tangent Kernel for PINNs, which is expressed as $$ 
K = \left[
\begin{array}{cc}
K_{uu} & K_{ur} \\
K_{ru} & K_{rr}
\end{array}
\right],
 $$
for:
 \begin{equation*}
     \begin{aligned}
 &(K_{uu})_{ij} =\big \langle \partial_{\boldsymbol{\zeta}} N({\boldsymbol{\zeta}},x_{b_i}), \partial_{\boldsymbol{\zeta}} N({\boldsymbol{\zeta}},x_{b_j})   \big \rangle ;\\
 &(K_{rr})_{ij} = \big \langle \partial_{\boldsymbol{\zeta}} R({\boldsymbol{\zeta}},x_{i}), \partial_{\boldsymbol{\zeta}} R({\boldsymbol{\zeta}},x_{j}) \big \rangle ; \\
 &(K_{ur})_{ij} = (K_{ru})_{ji} = \big \langle \partial_{\boldsymbol{\zeta}} N({\boldsymbol{\zeta}},x_{b_i}) ,  \partial_{\boldsymbol{\zeta}} R({\boldsymbol{\zeta}},x_{j}) \big \rangle,
     \end{aligned}
 \end{equation*}
where $R$ is the residual of the PDE, and $N({\boldsymbol{\zeta}}, x_b)$ is the value of $N$ on the boundary. 
 \subsection{NTK for the second and first-order formulations}
Using a similar approach \cite{NTK,NTK_PINN}, we develop the NTK for the second-order equation of the form 
$$\nabla \cdot \nabla \mu u =0, $$
and for the equivalent first-order formulation:
\begin{align*}
    &\nabla \mu u =v; \\
    &\nabla \cdot v =0.
\end{align*}
The loss function for a PINN that is trained with the second-order formulation, with strongly imposed boundary conditions, is
$$ L = \|R_2({\boldsymbol{\zeta}},x) \| = \frac{1}{2N}\sum_{i=1}^N R_2({\boldsymbol{\zeta}}, x_i)^2 ,$$
where $R_2({\boldsymbol{\zeta}},x) =\nabla \cdot \nabla \mu N({\boldsymbol{\zeta}},x)  $ and $N({\boldsymbol{\zeta}},x)$ is the PINN approximation of $u$.
For a gradient descent algorithm with a small step size $dt$, the network parameter ${\boldsymbol{\zeta}}$ varies according to $$\frac{d {\boldsymbol{\zeta}}}{dt} = -\nabla_{\boldsymbol{\zeta}} L({\boldsymbol{\zeta}},x) = \frac{-1}{N}\sum_{i=1}^N \partial_{\boldsymbol{\zeta}} R_2({\boldsymbol{\zeta}},x_i)  R_2({\boldsymbol{\zeta}},x_i).  $$
Next, we express the rate of change of $R_2$ during the gradient descent using the chain rule, to obtain
$$\frac{d R_2({\boldsymbol{\zeta}},x_j)}{dt} =  
\frac{d R_2({\boldsymbol{\zeta}},x_j)}{d\boldsymbol{\zeta}} \frac{d {\boldsymbol{\zeta}}}{dt} 
=
\frac{-1}{N}\sum_{i=1}^N \partial_{\boldsymbol{\zeta}} R_2({\boldsymbol{\zeta}},x_j)  \partial_{\boldsymbol{\zeta}} R_2({\boldsymbol{\zeta}},x_i)  R_2({\boldsymbol{\zeta}},x_i).$$ We then define the NTK for the second-order formulation as $$(K_2)_{ij} = \big \langle \partial_{\boldsymbol{\zeta}} R_2({\boldsymbol{\zeta}},x_j) , \partial_{\boldsymbol{\zeta}} R_2({\boldsymbol{\zeta}},x_i)    \big \rangle. $$

For the first-order formulation, with a strong imposition of boundary conditions, the loss function becomes $$ L= \|R_{1a}({\boldsymbol{\zeta}},x) \| + \|R_{1b}({\boldsymbol{\zeta}},x) \|= \frac{1}{2N}\sum_{i=1}^N R_{1a}{\boldsymbol{\zeta}}({\boldsymbol{\zeta}}, x_i)^2 + R_{1b}({\boldsymbol{\zeta}}, x_i)^2,$$
where $R_{1a} = \nabla (\mu N_1({\boldsymbol{\zeta}},x) ) -N_2({\boldsymbol{\zeta}},x) $, $R_{1b} = \nabla \cdot N_2({\boldsymbol{\zeta}},x) $, $N_1({\boldsymbol{\zeta}},x)$ is the PINN approximation of $u$, and $N_2({\boldsymbol{\zeta}},x)$ is the PINN approximation of $v$.

The rate of change of the parameter ${\boldsymbol{\zeta}}$ during the gradient descent is given by 
 $$\frac{d {\boldsymbol{\zeta}}}{dt} = -\nabla_{\boldsymbol{\zeta}} L({\boldsymbol{\zeta}},x) = \frac{-1}{N}\sum_{i=1}^N \partial_{\boldsymbol{\zeta}} R_{1a}({\boldsymbol{\zeta}},x_i)  R_{1a}({\boldsymbol{\zeta}},x_i) + \partial_{\boldsymbol{\zeta}} R_{1b}({\boldsymbol{\zeta}},x_i)  R_{1b}({\boldsymbol{\zeta}},x_i).  $$

The rate of change of $R_{1a}$ and $R_{1b}$ during the gradient descent is given by:
\begin{equation*}
\begin{aligned}
\frac{d R_{1a}({\boldsymbol{\zeta}},x_j)}{dt} =&  \frac{-1}{N}\sum_{i=1}^N  
\partial_{\boldsymbol{\zeta}} R_{1a}({\boldsymbol{\zeta}},x_j) \, \partial_{\boldsymbol{\zeta}} R_{1a}({\boldsymbol{\zeta}},x_i) \, R_{1a}({\boldsymbol{\zeta}},x_i)\\   
& + \partial_{\boldsymbol{\zeta}} R_{1a}({\boldsymbol{\zeta}},x_j) \, \partial_{\boldsymbol{\zeta}} R_{1b}({\boldsymbol{\zeta}},x_i) \, R_{1b}({\boldsymbol{\zeta}},x_i) ;
\end{aligned}
\end{equation*}
\begin{equation*}
\begin{aligned}
\frac{d R_{1b}({\boldsymbol{\zeta}},x_j)}{dt} =  &\frac{-1}{N}\sum_{i=1}^N 
\partial_{\boldsymbol{\zeta}} R_{1b}({\boldsymbol{\zeta}},x_j) \, \partial_{\boldsymbol{\zeta}} R_{1a}({\boldsymbol{\zeta}},x_i) \, R_{1a}({\boldsymbol{\zeta}},x_i) \\ 
&+  \partial_{\boldsymbol{\zeta}} R_{1b}({\boldsymbol{\zeta}},x_j) \, \partial_{\boldsymbol{\zeta}} R_{1b}({\boldsymbol{\zeta}},x_i) \, R_{1b}({\boldsymbol{\zeta}},x_i) .
\end{aligned}
\end{equation*}
Hence, we define the NTK associated with the first-order formulation as 
 $$ 
K_1 = \left[
\begin{array}{cc}
K_{aa} & K_{ab} \\
K_{ba} & K_{bb}
\end{array}
\right],
 $$ where: 
 \begin{equation*}
     \begin{aligned}
         &(K_{aa})_{ij} =\big \langle \partial_{\boldsymbol{\zeta}} R_{1a}({\boldsymbol{\zeta}},x_{i}) , \partial_{\boldsymbol{\zeta}} R_{1a}({\boldsymbol{\zeta}},x_{j})\big \rangle ;\\      
         &(K_{bb})_{ij} = \big \langle \partial_{\boldsymbol{\zeta}} R_{1b}({\boldsymbol{\zeta}},x_{i}) , \partial_{\boldsymbol{\zeta}} R_{1b}({\boldsymbol{\zeta}},x_{j}) \big \rangle ;\\     
         &(K_{ab})_{ij} = (K_{ba})_{ji} = \big \langle \partial_{\boldsymbol{\zeta}} R_{1a}({\boldsymbol{\zeta}},x_{i}) ,  \partial_{\boldsymbol{\zeta}} R_{1b}({\boldsymbol{\zeta}},x_{j})\big \rangle.
     \end{aligned}
 \end{equation*}
 
 In both cases, the evolution of the residual is expressed as $$\frac{dR}{dt} = -K  R,$$ where $R = R_2$ for the second-order formulation, and $R = [R_{1a}, R_{1b}]$ for the first-order formulation.
 
 Since $K$ is a symmetric semi-definite matrix, it can again be expressed as $K = Q D Q^T$, leading to a solution of the form 
 $$Q^T R = e^{-Dt} Q^T R_{(t=0)}. $$

If we consider the example of a second-order formulation with $\mu = S (100(0.5-x) ),$ where $S$ is the sigmoid function, the kernel $K_2$ becomes
\begin{align*}  
(K_2)_{ij} &= \big \langle \partial_{\boldsymbol{\zeta}} (\mu''  N({\boldsymbol{\zeta}},x_i) + 2\mu'  N'({\boldsymbol{\zeta}},x_i) +  \mu N''({\boldsymbol{\zeta}},x_i) ) , \\ 
& \qquad \partial_{\boldsymbol{\zeta}}( \mu''  N({\boldsymbol{\zeta}},x_j) + 2\mu'  N'({\boldsymbol{\zeta}},x_j) +  \mu N''({\boldsymbol{\zeta}},x_j) )  \big \rangle \\
&\approx 
\big \langle \mu''  
\partial_{\boldsymbol{\zeta}} ( N({\boldsymbol{\zeta}},x_i)) ,
\mu''  
\partial_{\boldsymbol{\zeta}}( N({\boldsymbol{\zeta}},x_j)) \big \rangle,
\end{align*}
because the magnitude of $\mu''$ is dominant. Since $\mu''$ is large around $x=0.5$, and zero elsewhere, we can expect the eigenvectors of $K_2$ to be dominant around $x=0.5.$
On the other hand, the loss function for the first-order formulation, $$L = \|\partial_x N_2 \| + \| \partial_x(\mu N1) - N_2 \|, $$ still with $\mu = S (100(0.5-x) )$,  is dominated by the term $$\| N_2({\boldsymbol{\zeta}},x)) - \partial_x ( \mu N_1({\boldsymbol{\zeta}},x)) \| \approx \|\partial_x ( \mu N_1({\boldsymbol{\zeta}},x) ) \|,$$
so we can approximate $$K_{aa} \approx 
\big \langle \mu'  
\partial_{\boldsymbol{\zeta}} ( N_1({\boldsymbol{\zeta}},x)) + \mu \partial_{\boldsymbol{\zeta}} ( N_1'({\boldsymbol{\zeta}},x)),
\mu'
\partial_{\boldsymbol{\zeta}}( N_1({\boldsymbol{\zeta}},x)) + \mu \partial_{\boldsymbol{\zeta}} ( N_1'({\boldsymbol{\zeta}},x)) \big \rangle,$$ and since $\max(\mu')\approx 25$ and $\max(\mu'') \approx 1000$, we can expect the first-order formulation to be less restrictive than the second-order formulation at $x=0.5.$

\subsection{Numerical study of the effect of the formulation on eigenvectors, eigenvalues, and on the convergence properties of the PINN}
 
We conduct two numerical tests to study the effect of the equations' formulation on $Q$, $D$, and $R_{t=0}$.
 In the first numerical test, we consider the 1D second-order problem
 $$ \nabla \cdot \nabla \mu  u = 0, \ \text{where} \  u(0)=1 \text{and} \ u(1)=0 .$$ 
The equivalent first-order formulation of the problem is:
\begin{align*}
 &v = \nabla \mu  u; \\
 &\nabla \cdot v =0, \ \text{where} \  u(0)=1 \text{and} \ u(1)=0. 
 \end{align*}
We will compare the results obtained for $\mu = S (100(0.5-x)),$ and for $\mu =1$.

% At initialization, where $v$ and $u$ are two random and smooth functions, the loss function, and subsequently, the kernel of the first-order formulation, will be dominated by the term $$ \|v -  \nabla \mu  U\| \approx \| \nabla \mu  U\|,$$ so for simplification and ease of visual representation, we will compare the eigenvalues and eigenvectors of the kernels obtained by $R_1 = \|\nabla \mu U \|$ and $R_2 = \| \nabla \cdot \nabla \mu U \|$.

\begin{figure}[h]
\begin{subfigure}[t]{0.5\textwidth}
\centering
    \includegraphics[width=8cm]{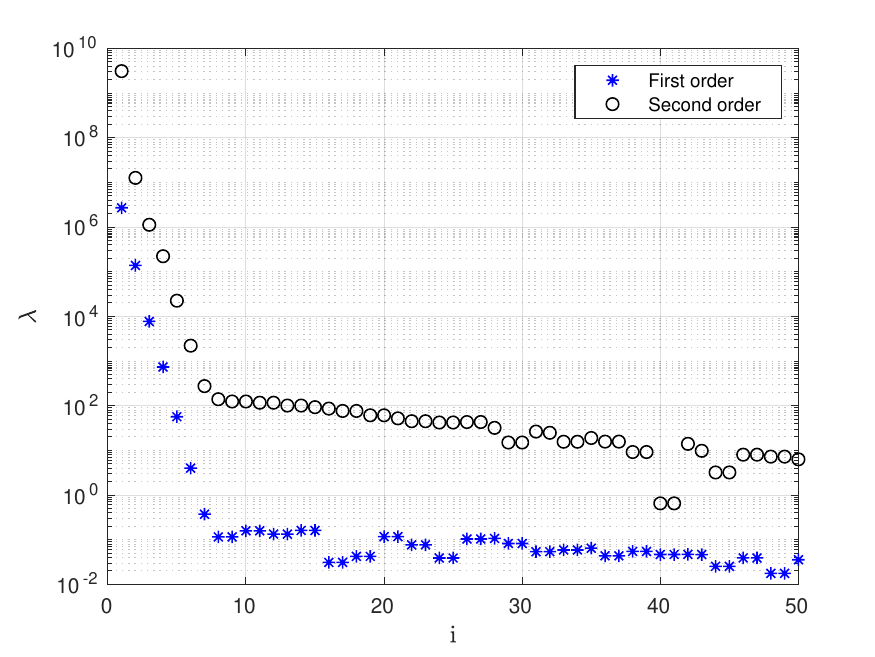}
\caption{}
\end{subfigure}
\begin{subfigure}[t]{0.5\textwidth}
\centering
    \includegraphics[width=8cm]{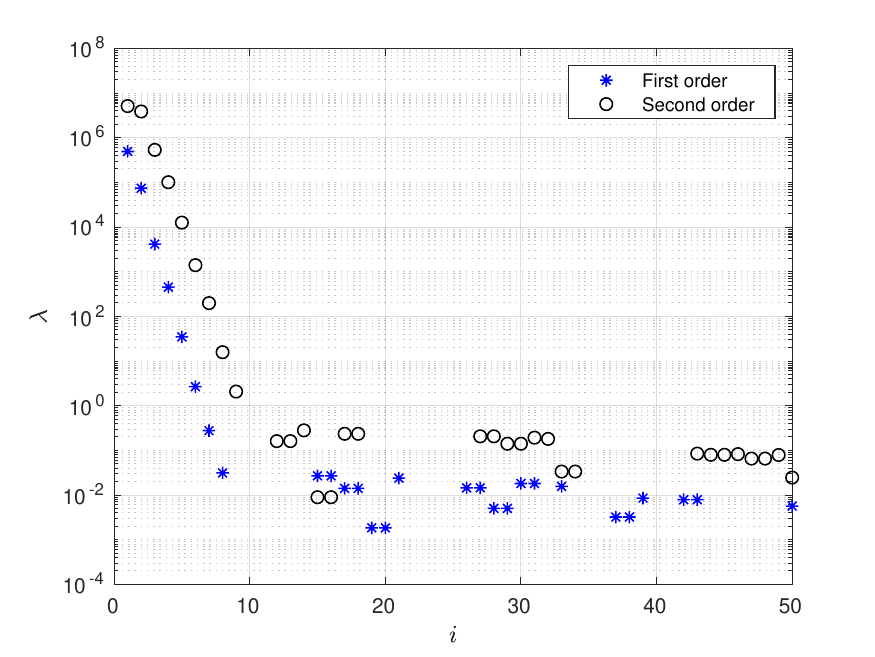}
\caption{}
\end{subfigure}
    \caption{ The first 50 eigenvalues of $K_1$ and $K_2$ for the first NTK numerical test, obtained using the first and second-order formulations for (a) \mbox{$\mu = S ( 100(0.5-x) ) $ }, (b) $\mu =1 $.}
    \label{fig:lambda}
\end{figure}

First, we compare the computed eigenvalues of $K_1$ and $K_2$, illustrated in figure~\ref{fig:lambda}. We notice the difference in magnitude, portrayed in Table~\ref{tab:lambda}, between the leading eigenvalues of $K_2$ for $\mu = S ( 100(0.5-x) )$, compared to the difference in magnitude between the leading eigenvalues of $K_1$. We also notice that we do not have a difference in magnitude for the case of $\mu=1$.

\begin{table}[ht]
    \centering
    \begin{tabular}{|c|c|c|c|c|} \hline 
          order&&  $\frac{\lambda_1}{\lambda_2}$&  $\frac{\lambda_2}{\lambda_3}$&  $\frac{\lambda_3}{\lambda_4}$\\ \hline 
          2&$\mu_1$ &  141&  5&  2\\ \hline 
          1&$\mu_1$&  3&  5&  2\\ \hline 
          2&$\mu_2$&  2&  7&  5\\ \hline 
          1&$\mu_2$&  7&  4&  4\\ \hline
    \end{tabular}
    \caption{The difference in magnitude between the leading eigenvalues, using $\mu_1 =S ( 100(0.5-x) ) $ and $\mu_2=1$, with the second-order formulation and the first-order formulation, for the first NTK numerical test.}
    \label{tab:lambda}
\end{table}

\begin{figure}[ht]
\centering
    \includegraphics[width=6.5cm]{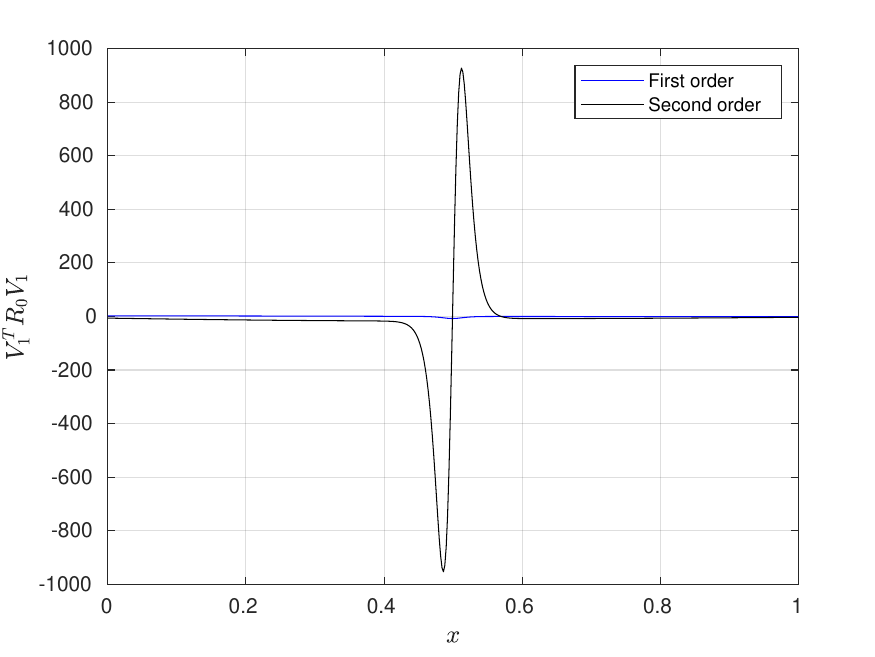}
    \includegraphics[width=6.5cm]{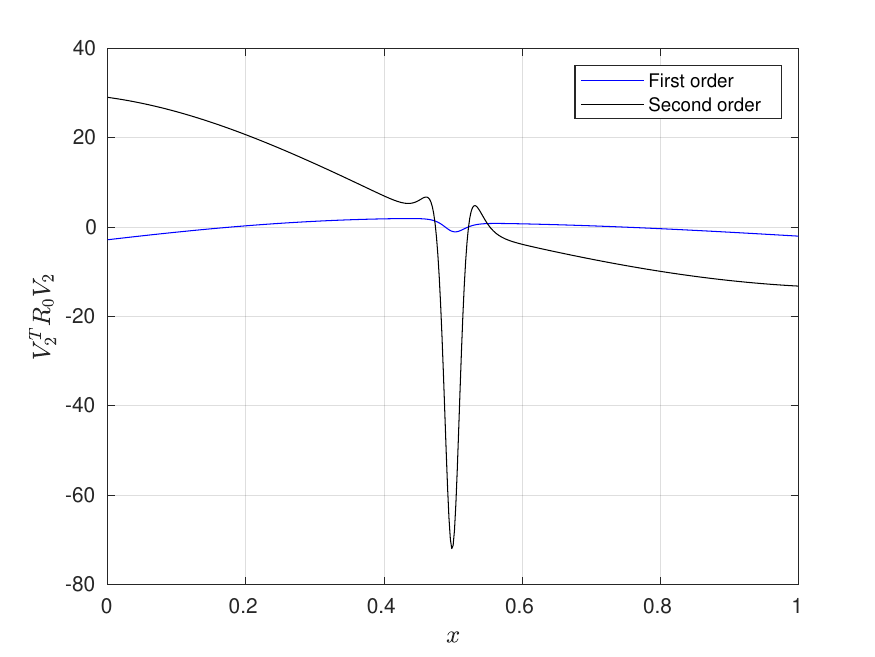}
    \includegraphics[width=6.5cm]{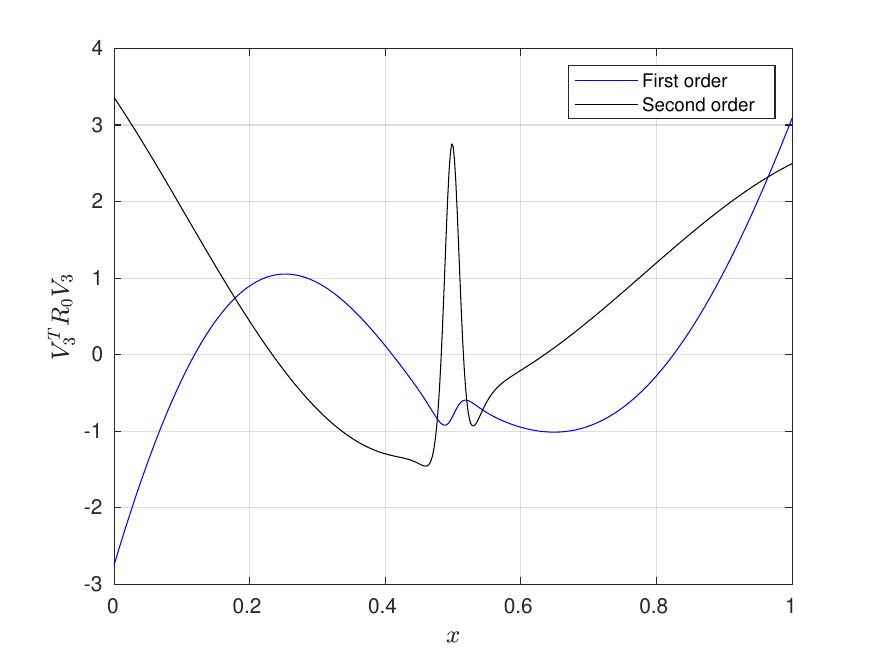}
    \includegraphics[width=6.5cm]{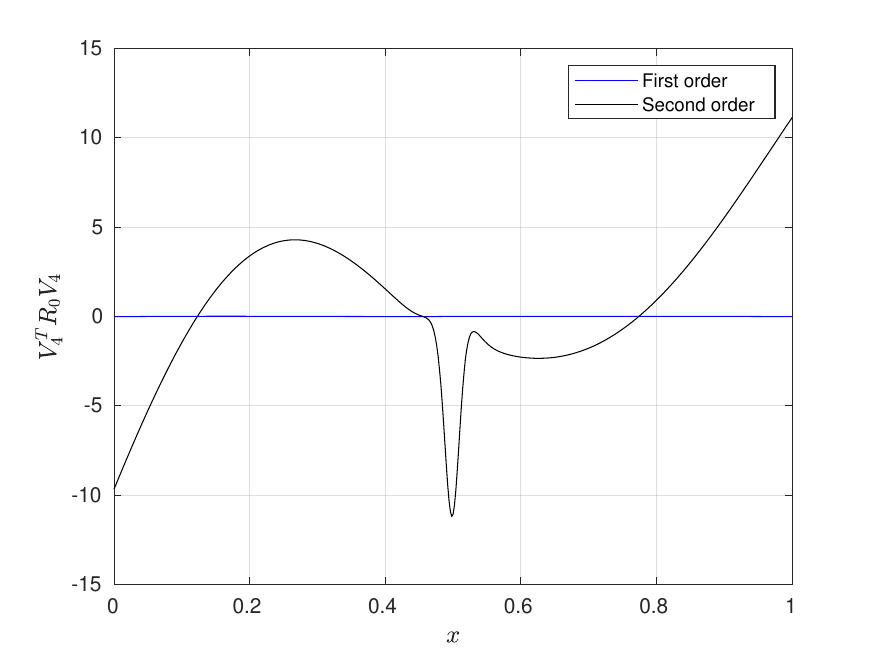}
    \caption{ The projection of the residuals on the first four eigenvectors for the first NTK numerical test with \mbox{$\mu = S ( 100(0.5-x) ),$}.  }
    \label{fig:QTR}
\end{figure}

Next, we compare the projection of the residuals on the eigenvectors of the two kernels, $$P = Q^T R_0,$$ illustrated in Figure~\ref{fig:QTR}. We notice the difference in magnitude between the projections of the first-order and the second-order formulations, and that the eigenvectors of the second-order formulation are large around $x=0.5$, when compared to the rest of the domain.

 \begin{figure}[ht]
 \centering
    \includegraphics[width=6.5cm]{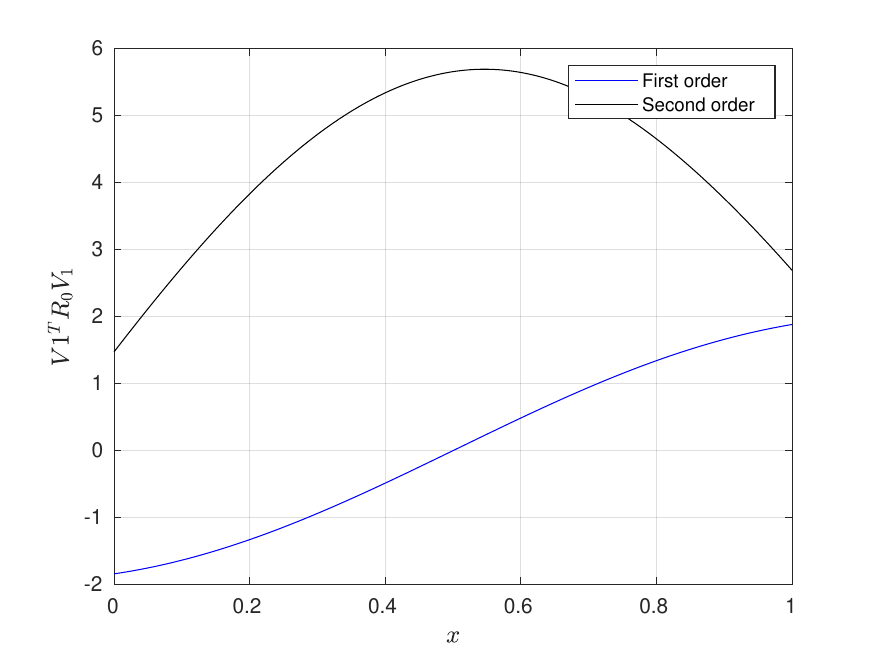}
    \includegraphics[width=6.5cm]{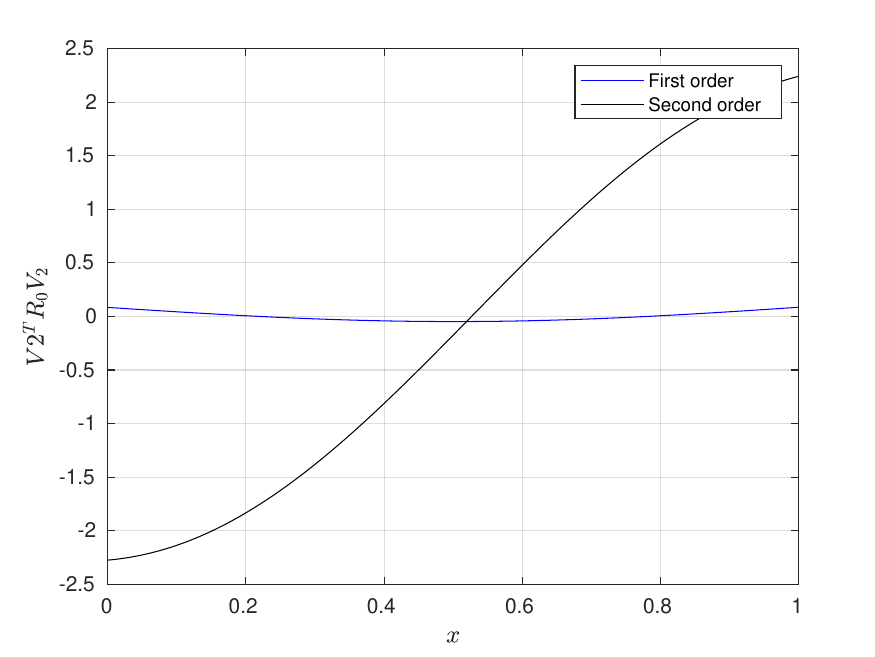}
    \includegraphics[width=6.5cm]{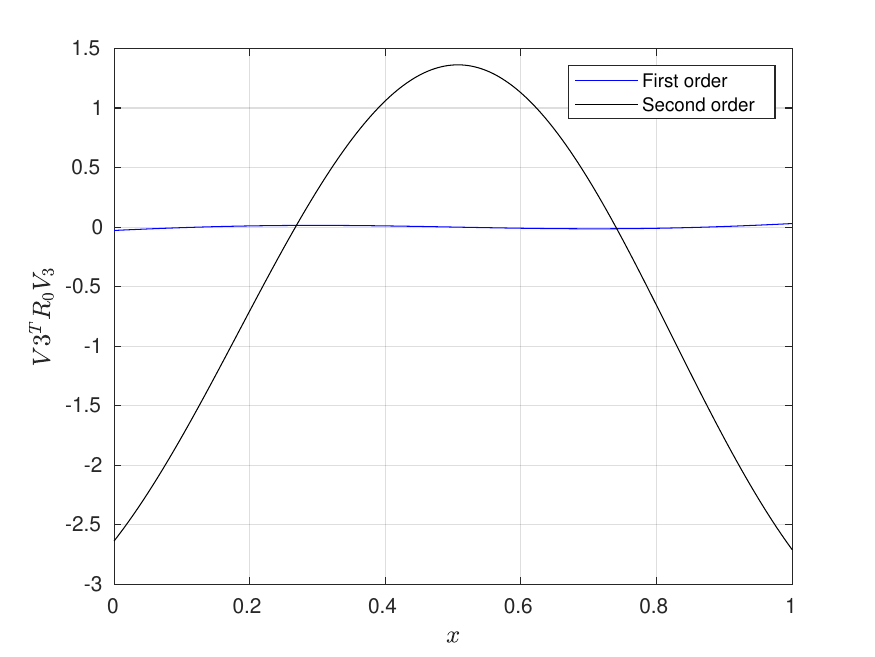}
    \includegraphics[width=6.5cm]{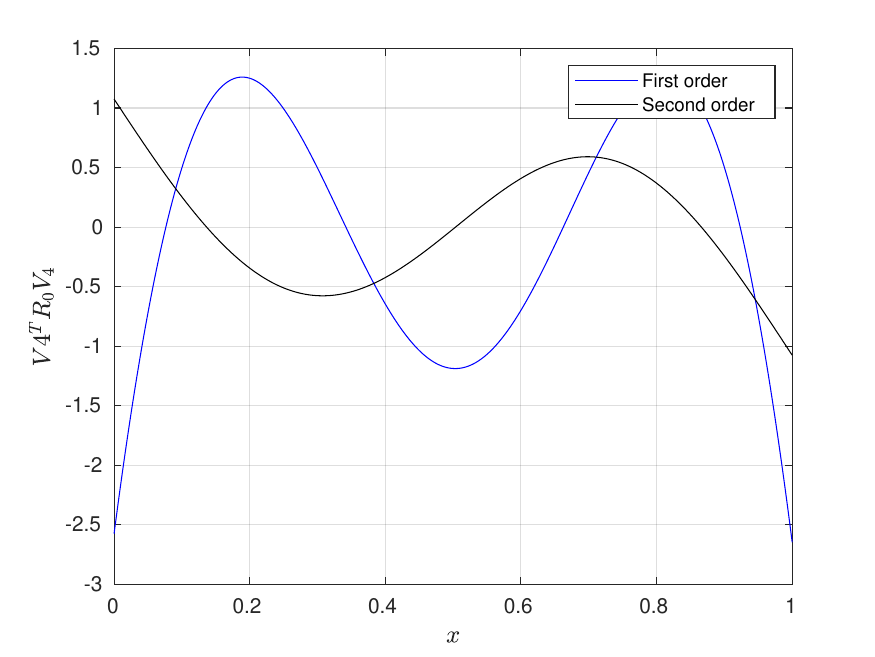}
    \caption{ The projection of the residuals on the first four eigenvectors for $\mu = 1$, for the first NTK numerical test.}
    \label{fig:QTR2}
    
\end{figure}

The difference in magnitude between the eigenvalues of $K_2$ suggests that the network will prioritize minimizing the residual along the first eigenvector, which will stiffen the learning process. The large magnitude of the projection over the remaining eigenvectors indicates that the network will converge to a local minimum, or plateau, with large residuals.

\begin{figure}[h]
\begin{subfigure}[t]{0.5\textwidth}
\centering
    \includegraphics[width=7.5cm]{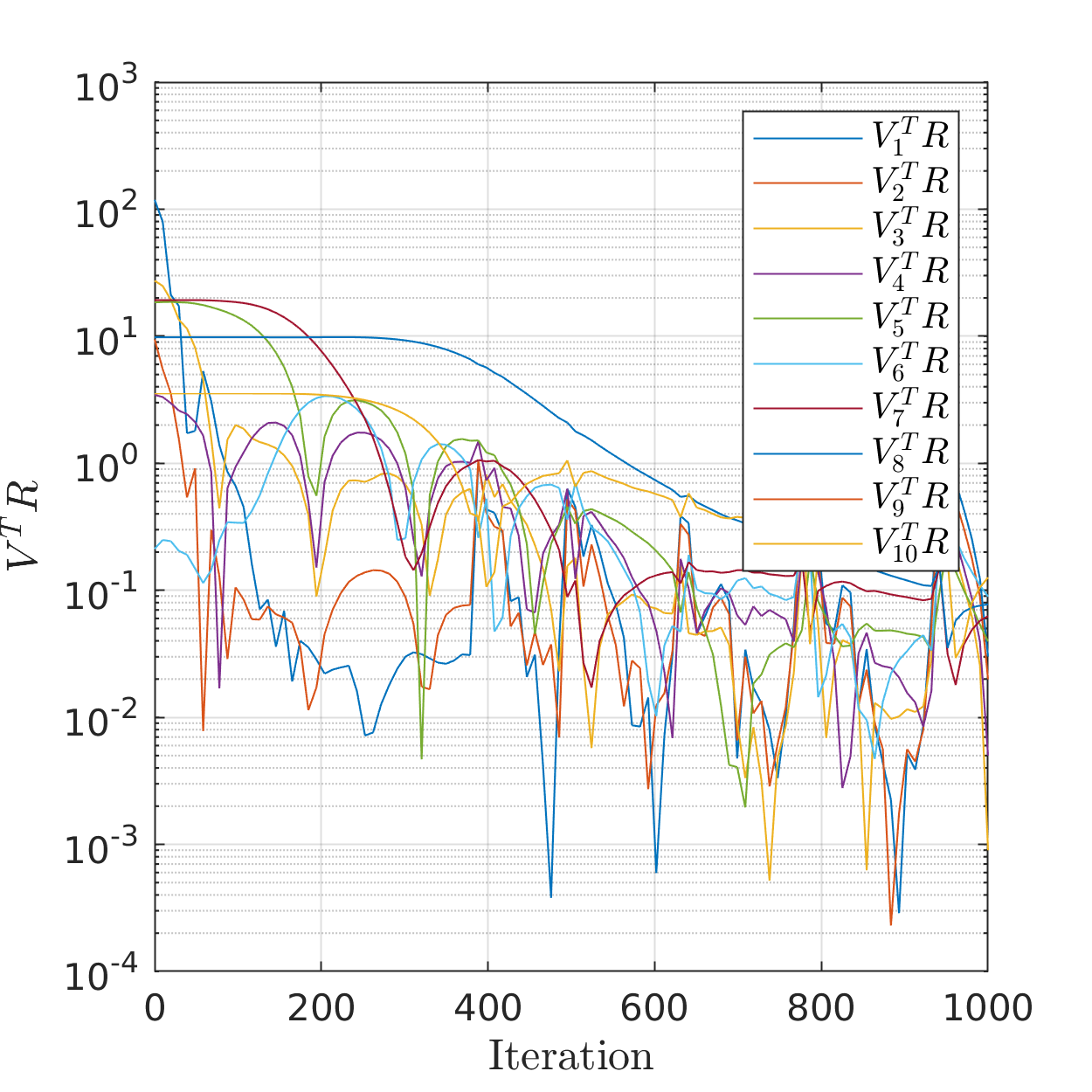}
    \caption{}
\end{subfigure}
\begin{subfigure}[t]{0.5\textwidth}
    \includegraphics[width=7.5cm]{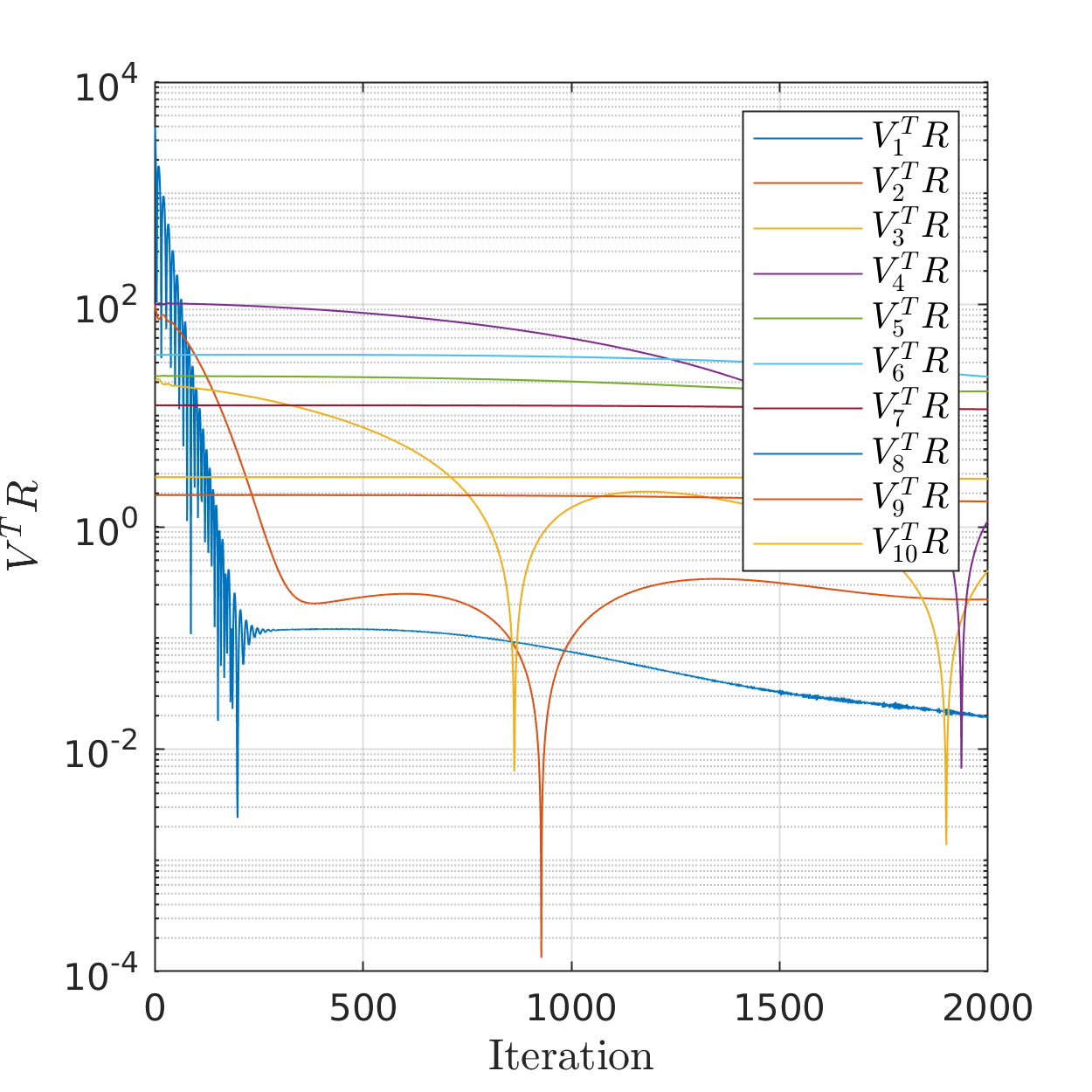}
    \caption{}
\end{subfigure}
    \caption{ The magnitude of the projection of the residuals on the first ten eigenvectors during training for (a) the first-order formulation, (b) the second-order formulation, for the first NTK numerical test with $\mu = S ( 100(0.5-x))$.}
    \label{fig:QTR_a}
\end{figure}

\begin{figure}[ht]
\begin{subfigure}[t]{0.3\textwidth}
\centering
    \includegraphics[width=\linewidth]{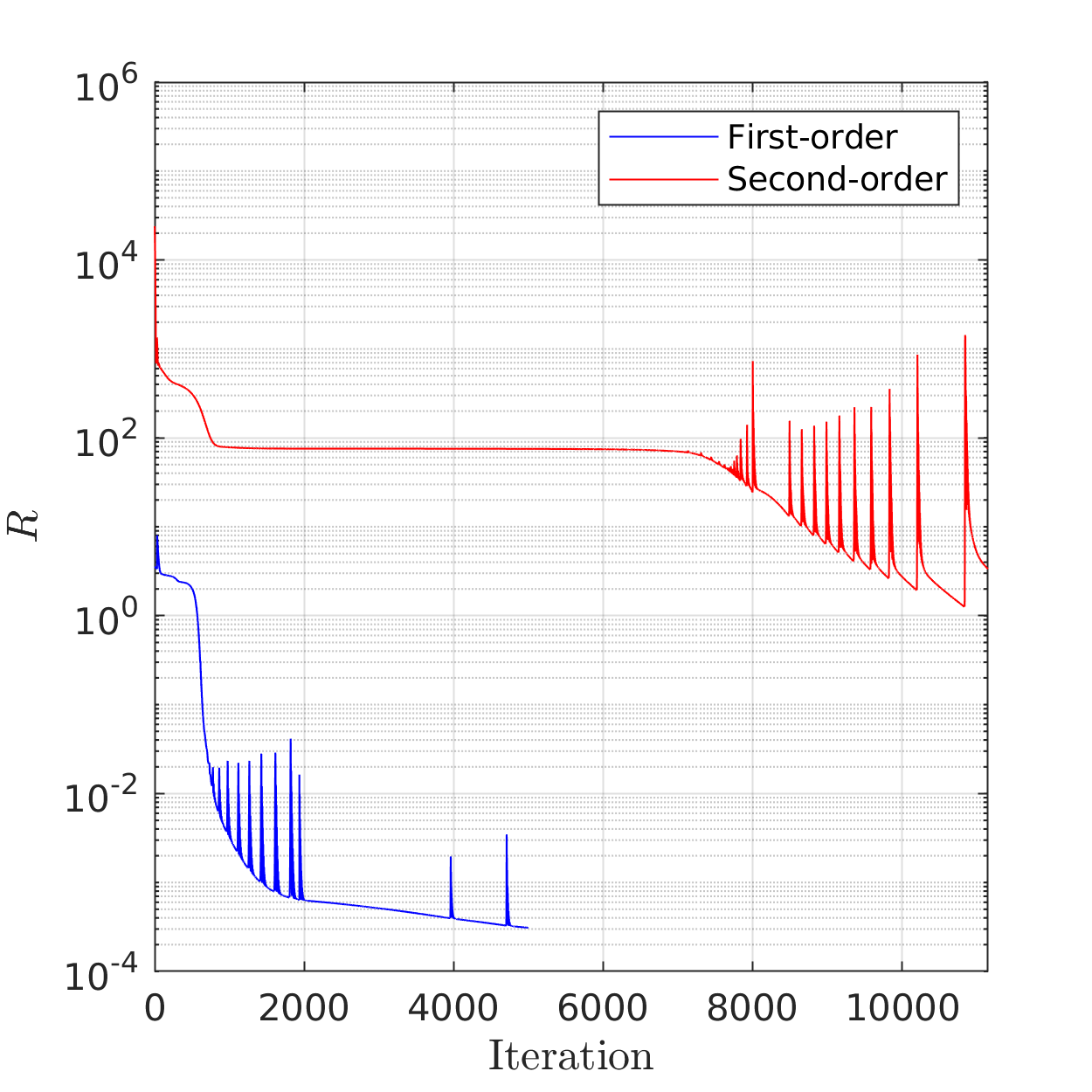}
    \caption{}
\end{subfigure}
\begin{subfigure}[t]{0.3\textwidth}
    \includegraphics[width=\linewidth]{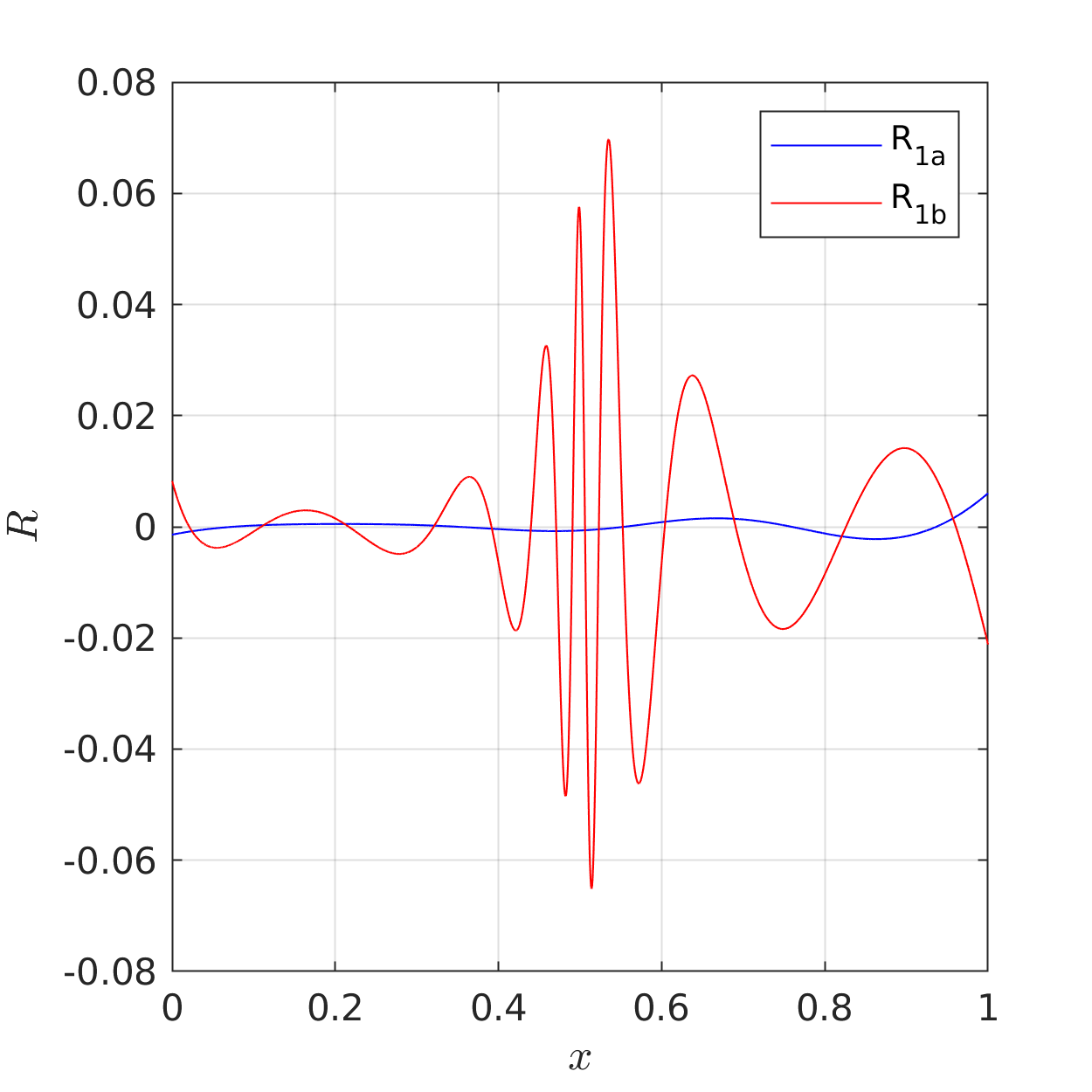}
    \caption{}
\end{subfigure}
\begin{subfigure}[t]{0.3\textwidth}
    \includegraphics[width=\linewidth]{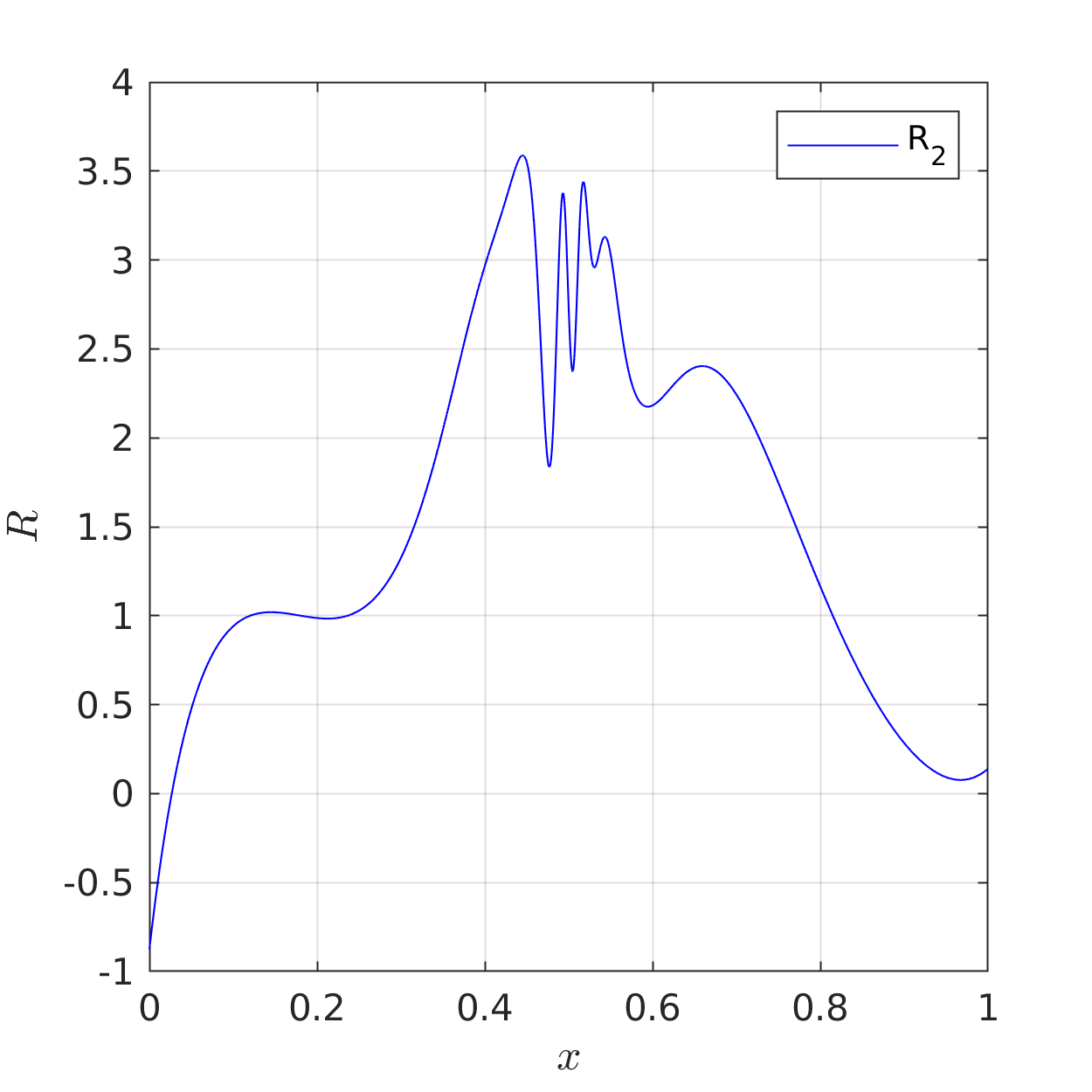}
    \caption{}
\end{subfigure}
    \caption{(a) The evolution of the residuals obtained by the two formulations during training, for the first NTK numerical test. (b) The residuals at the final iteration $n=5000$ for the first-order formulation. (c) The residual at the final iteration $n=11 000$ for the second-order formulation.}
    \label{fig:fo2o_L}
\end{figure}

The conclusion reached above can be verified by looking at figure~\ref{fig:QTR_a}, which shows the evolution of the projections on the eigenvectors, $Q^T R,$ during training. We see that for the first-order formulation, the projections vary synchronously, and the network can explore weights that result in larger projections on the first eigenvector but smaller overall residuals. In contrast, with the second-order formulation, the projection on the first eigenvector diminishes quickly and remains low, while the network struggles to converge over the other projections, which leads to a stiff convergence.
% In Figure~\ref{fig:fo2o_e2}, we show that the residual of the first-order formulation is indeed dominated by the term $\| \nabla \mu  U\|$, and 
In figure~\ref{fig:fo2o_L}, we see that the second-order formulation hits a plateau for a large residual, for an extended training period, while the first-order formulation quickly converges.

%  \begin{figure}[ht]
% \centering
%     \includegraphics[width=7.5cm]{exp2_o1.pdf}
%     \includegraphics[width=7.5cm]{exp2_o2.pdf}
%     % \includegraphics[width=7.5cm]{fft1.pdf}
%     % \includegraphics[width=7.5cm]{fft2.pdf}
%     \caption{ The initial residual obtained by the first and second-order formulation.}
%     \label{fig:fo2o_e2}
% \end{figure}

For the second numerical test, we consider the 3D second-order boundary value problem $$ \nabla \cdot
 \nabla U = F_{\rm ext},$$
 where the domain is a unit cube, with homogeneous Dirichlet boundary conditions on all sides of the cube. The analytical solution is 
 $$U = \cos(5\pi x)\cos(5\pi y)\cos(5\pi z).$$
The first-order formulation of the problem is:
 \begin{equation*}
 \begin{split}
\nabla U = \boldsymbol{V}; \\
\nabla \cdot \boldsymbol{V} = F_{\rm ext},
\end{split}
  \end{equation*}
where we introduce three auxiliary variables (for the three-dimensional case).

The boundary conditions are strongly enforced. The loss function of the second-order formulation is $$ L = \|\nabla \cdot \nabla U - F_{\rm ext} \| ,$$
and the loss function for the first-order formulation is  $$L = \beta \| \nabla \mathbf{U} - V\| +\|\nabla \cdot \mathbf{V} - F_{\rm ext} \| ,$$
where $\beta$ is a weight parameter.

\begin{figure}
\centering
    \includegraphics[width=0.7\linewidth]{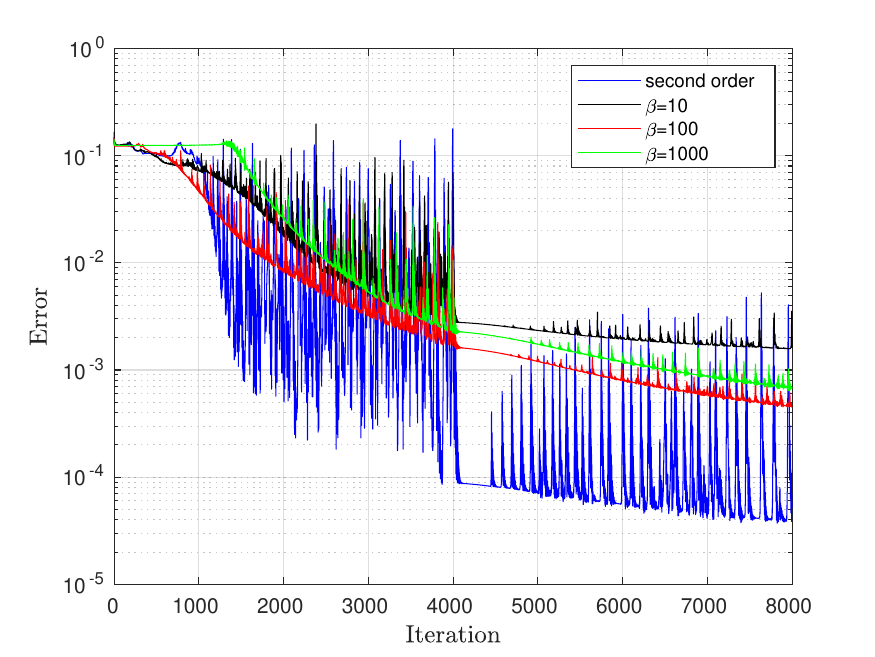}
    \caption{ The variation of the error per iteration for various values of the weight parameter $\beta$, with the first-order formulation, and the second-order formulation, for the second NTK numerical test.}
    \label{fig:fo2o_e1}
\end{figure}
From the results illustrated in figure~\ref{fig:fo2o_e1}, we can see that the second-order formulation yields more accurate results for the same number of iterations. More so, we can see the dependency of the approximation's accuracy on $\beta$ in the first-order formulation. For a small value of $\beta$, the discrepancy between $\nabla
 U$ and $\mathbf{V}$ leads to inaccurate results, and for large values of $\beta$, the contribution of $\|\nabla \cdot \mathbf{V} - F_{\rm ext} \|$ to the loss function is negligible.

From these findings, we conclude that the first-order formulations are better suited for interface problems, and we will be using the first-order formulation of Maxwell's equations \eqref{eqn:lowf_max_2}.

\section{Numerical study on the effect of the hyperparameters on the PINN with strongly imposed boundary conditions}

With the vanilla PINN~\cite{raissi}, the network is trained by minimizing a loss function that mainly contains the residuals of the governing PDEs, the boundary conditions, and known values of the solution. The large number of objectives in the loss function can result in a stiff convergence of the PINN. The usual remedy is to add weights to each term in the loss function. Multiple studies were conducted to determine the optimal values for these weights~\cite{NTK_PINN,bischof2021multi,maddu2022inverse,deguchi2023dynamic}. However, in this paper, we will use a PINN with strong imposition of boundary and initial conditions.
We do so to avoid boundary errors, and the arbitrary choice of weights in the loss function. 
The architecture of a PINN with the strong imposition of boundary conditions is illustrated in figure~\ref{fig:net_arch_T2}, where $d(\x)$ is a function that is zero on the boundary, such as the distance function to the interface, and $V(\x)$ is a lift function that satisfies the boundary conditions.

\begin{figure}
% \begin{center}

% \tikzstyle{block} = [draw, fill=white, rectangle, 
%     minimum height=3em, minimum width=6em]
% \tikzstyle{sum} = [draw, fill=white, circle, node distance=1cm]
% \tikzstyle{input} = [coordinate]
% \tikzstyle{output} = [coordinate]
% \tikzstyle{pinstyle} = [pin edge={to-,thick,black}]

% \begin{tikzpicture}[auto, node distance=2cm,>=latex]

%     \node [block, name=input] (controller) {$\boldsymbol{x}$ };
    
%     \node [block, right of=controller,
%             node distance=4cm] (system) {Neural Network $N(\boldsymbol{x})$};
            
%     \node [block, above of=system] (boundary) {boundary function $V(\boldsymbol{x})$};
%     \node [block, below of=system] (distance) {distance function $d(\boldsymbol{x})$};
    
%     \node [block, right of=system,node distance=5cm] (sol) {$f(\boldsymbol{x})=N(\boldsymbol{x})d(\boldsymbol{x}) + V(\boldsymbol{x})$};

%     \draw [->,very thick] (controller) -- node[name=u] {} (boundary);
%     \draw [->,very thick] (controller) -- node[name=u] {} (system);
%     \draw [->,very thick] (controller) -- node[name=u] {} (distance);
    
%     \draw [->,very thick] (system) -- node[name=u] {} (sol);
%     \draw [->,very thick] (boundary) -- node[name=u] {} (sol);
%     \draw [->,very thick] (distance) -- node[name=u] {} (sol);
  
% \end{tikzpicture}

%     \centering
    
%     \caption{The architecture of the neural network with the strong imposition of boundary conditions.}
%     \label{fig:net_arch_M}
% \end{center}
\end{figure}

To determine the effect of the functions $d(\x)$ and $V(\x)$ on the convergence and accuracy of the approximated solution, we will conduct a test on the system of first-order differential equations:
\begin{align*}
    &u' + v=30\cos{30 x}; \\
    &u - v' =0; \\
    &u(0) = 1, \ v(0) = 0.
\end{align*}
The analytical solution of this problem is given by
$$v= \sin(x) + \dfrac{30}{899}\cos(x) -  \dfrac{30}{899}\cos(30x),$$
and $$u =  \cos(x) - \dfrac{30}{899}\sin(x) +  \dfrac{900}{899}\sin(30x).$$

\begin{figure}
    \centering
    \includegraphics[width=8cm]{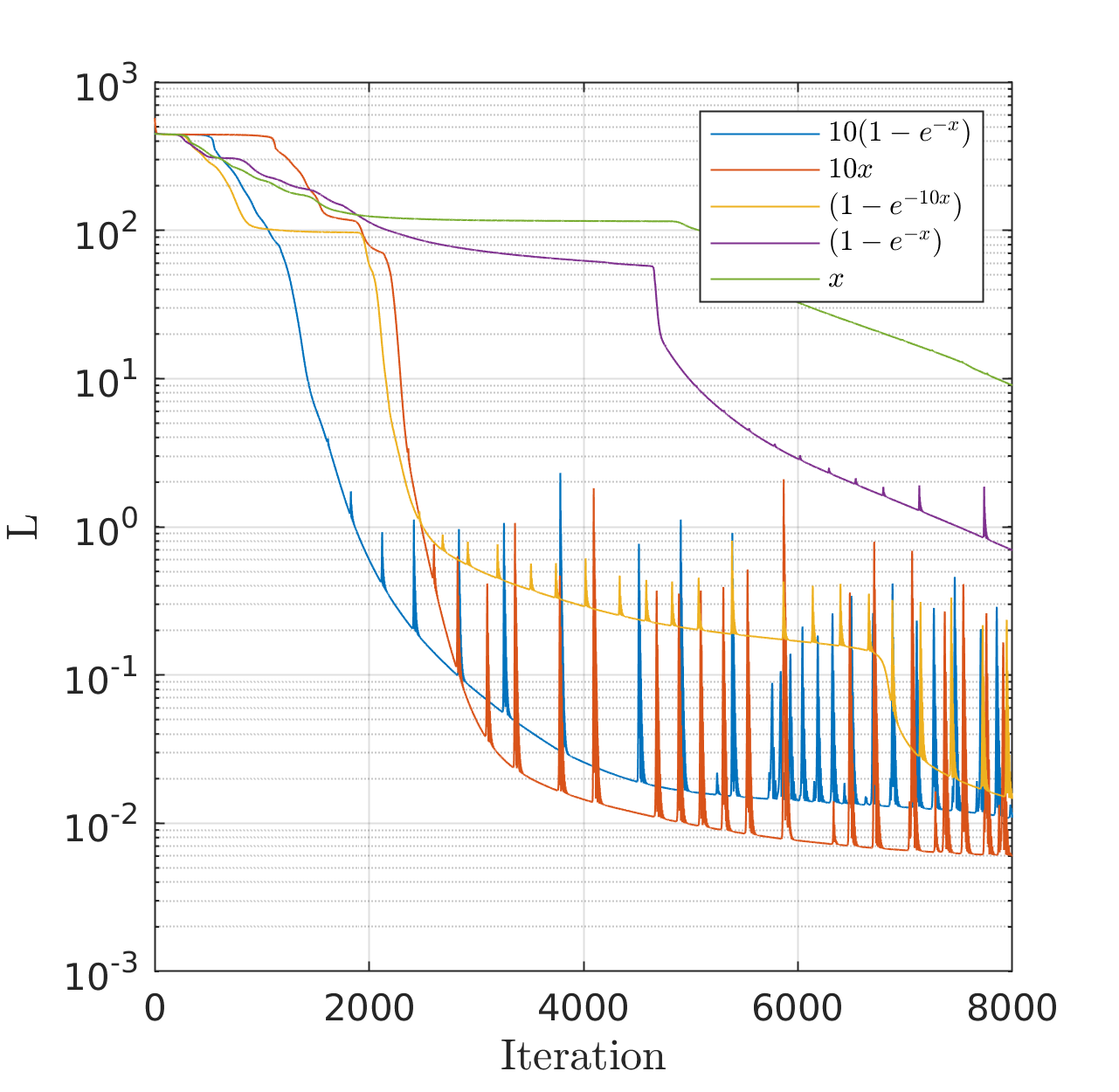}
    \caption{Evolution of the loss function during training for different distance functions $d(\x)$ for the numerical study on the effect of the hyperparameters on PINNs with strongly imposed boundary conditions.}
    \label{fig:conv_d(x)}
    
\end{figure}

The solutions $u$ and $v$ are approximated with a PINN, with a strong imposition of the boundary condition of the form:
\begin{align*}
    &u = N_1({\boldsymbol{\zeta}},x) \, d(\x) +1; \\
    &v=N_2({\boldsymbol{\zeta}},x) \, d(\x),
\end{align*}
where $N_1({\boldsymbol{\zeta}},x)$ and $N_2({\boldsymbol{\zeta}},x)$ are the outputs of the neural networks.
 The test is performed for various functions $d(\x),$ as illustrated in figure~\ref{fig:conv_d(x)}.
From these results, we can see that the function $d(\x)$ plays an important role in the convergence of PINNs, and that the functions $d(\x)$ that showed promising results all have a derivative at the boundary with the same order of magnitude as the derivative of the exact solution. We explain this behavior by denoting that the approximate solution at the boundary is 
\[u'(0)= N_1'(0)  d(0) + d'(0)  N_1(0)=d'(0)  N_1(0).\]
Hence, the value of $N_1(0)$ is inversely proportional to the value of $d'(0)$, and for faster convergence, the output of the neural network needs to be normalized, with an order of magnitude of 1.
Thus, by taking the function $d(\x)$ with a derivative on the boundary with the same order of magnitude as the derivative of the analytical solution, the function $N_1(\x)$ would have a magnitude close to 1. \\
We did not find any clear relation between the choice of the function $V(x)$ and the convergence speed. The only constraint on $V(x)$ is for the function to be smooth.

% \subsubsection{convergence analysis}
% In this subsection, we intend to demonstrate that the predicted solution by our method is bounded by the loss function. Meaning that as the norm of the loss function approaches zero, the predicted solution approach the exact solution.
% consider the equation $L(u)=f$ where $L$ is a linear differential operator, $u$ is the exact solution and $f$ is the right hand side. The approximated solution $u_h$ by PINN could then be written as $u_h=u+e$. By substitution we obtain $$L(u_h)-f=Loss$$
% $$L(u-e)-f=L(u)-L(e)-f=L(e)=Loss $$
% Since we strongly imposing the boundary condition on $u_h$, we obtain that the boundary conditions for $e$ is null, and the only solution for $L(e)=0$ is $e=0$.
% considering that the differential operator $L$ equipped with the right boundary condition is bijective and thus invertible, we obtain that $$e=L^{-1}(Loss) $$
% $$ \|L^{-1}(Loss) \| \le c \|L^{-1}\| \cdot \|Loss\| $$
% $$ \|e \| \le c \|L^{-1}\| \cdot \|Loss\|. $$
% This proves that the convergence toward the solution is of first-order with respect to the loss function.

% \begin{figure}
%     \centering
%     \includegraphics[width=8cm]{conv.png}
%     \caption{norm of the error with respect to the norm of the loss function}
    
% \end{figure}
\section{Proposed methodology}
\label{sec:method}
To better understand the challenges related to the modeling of material interfaces using PINNs, we will study a 1D example before discussing the methodology for addressing this issue.
The boundary-value problem is given by
$$\nabla \cdot ( \mu \, v) = 0,\ \text{with}\ v(0) = 1,$$ where $\mu$ is piecewise constant,

$$\mu=
\begin{cases} 
      1, & x\leq 0.5 ;\\
      2, &  x>0.5 ,
\end{cases}
$$

and the analytical solution is

$$v=
\begin{cases} 
      1 ,& x\leq 0.5 ;\\
      0.5, & x>0.5.
   \end{cases}
$$

\begin{figure}[h]
    \centering

    \begin{tikzpicture}[>=latex, y=0.5cm, font=\small]
        \draw[thick,->] (0,0) -- (4.5,0) node[anchor=north west] {$\textit{x}$};;
        \draw[thick,->] (0,0) -- (0,4.5) node[anchor=north west] {$\mu$};;
        \draw (0,1) -- (2,1);
        \draw (2,1) -- (2,2);
        \draw (2,2) -- (4,2) node[anchor=north west] {$\mu$};;
    
        \foreach \Point in { (0,0), (0.7,0), (1.4,0), (2.4,0), (3,0), (3.8,0) }{
        \node at \Point {\textbullet};
        }
      \end{tikzpicture}
    \caption{Discontinuous material interface $\mu$ in 1D, and the collocation points along the $x$-axis, for the 1D example of section \ref{sec:method}.}
    \label{fig:1D_mu_ex}
\end{figure}
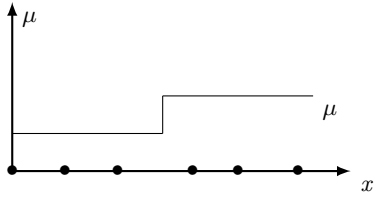

 If we attempt to train the PINN with the point distribution illustrated in Figure~\ref{fig:1D_mu_ex}, we could obtain that $v=1$, since the equation 
 $$\nabla \cdot (\mu v) =v \cdot \nabla \mu + \mu  \nabla \cdot v= 0$$ 
 is satisfied at each collocation point, and the boundary condition is satisfied at $x=0.$
The usual approach is to add collocation points at the interface, and include the interface jump condition in the loss function. 

In this work, we use a smooth approximation of the Heaviside function, calculated from the signed distance function $\phi(\x)$ to the interface, $$\hat{H} = S(\alpha \phi(\x)) ,$$ where $S$ is the sigmoid function, and $\alpha$ is a user-defined parameter, to approximate the physical quantities:

    $$ \mu = \mu_1 (1-\hat{H}) + \hat{H} \mu_2 $$
    and
    $$ \sigma = \sigma_1 (1-\hat{H}) + \hat{H} \sigma_2.$$

With this approach, we only need to add collocation points near the interface, and the interface jump conditions are not needed in the loss function.

\begin{figure}

\tikzstyle{block} = [draw, fill=white, rectangle, 
    minimum height=3em, minimum width=6em]
\tikzstyle{sum} = [draw, fill=white, circle, node distance=1cm]
\tikzstyle{input} = [coordinate]
\tikzstyle{output} = [coordinate]
\tikzstyle{pinstyle} = [pin edge={to-,thick,black}]

\begin{tikzpicture}[auto, node distance=2cm,>=latex]

    \node [block, name=input] (controller) {$\boldsymbol{x}$ };
    
    \node [block, right of=controller,
            node distance=4cm] (system) {common block $\boldsymbol{z}=N(\boldsymbol{x})$};
            
    \node [block, above right of=system, node distance=3cm] (FF) { $\boldsymbol{f}_1(\boldsymbol{z})$};
    \node [block, below right of=system, node distance=3cm] (SF) { $\boldsymbol{f}_2(\boldsymbol{z})$};
    
    \node [block, right of=system,node distance=5cm] (sol) {$\boldsymbol{f}=\boldsymbol{f}_1 + \hat{H} \boldsymbol{f}_2 $};
%%%%%%%%%%%%%%%%%%%%%%
    \node [block, above of=FF] (boundary) { $\boldsymbol{V}(\boldsymbol{x})$};
    \node [block, below of=SF] (distance) {$\boldsymbol{d}_0(\boldsymbol{x})$};
     \node [block, right of=sol,node distance=5cm] (Final_sol) { $\boldsymbol{y} = \boldsymbol{f}(\boldsymbol{x}) \, d_0(\boldsymbol{x}) + \boldsymbol{V}(\boldsymbol{x})$};
%%%%%%%%%%%%%%%%%%%%%%%%
    \draw [->,very thick] (system) -- node[name=u] {} (FF);
    \draw [->,very thick] (controller) -- node[name=u] {} (system);
    \draw [->,very thick] (system) -- node[name=u] {} (SF);
    \draw [->,very thick] (controller) -- node[name=u] {} (distance);
    \draw [->,very thick] (controller) -- node[name=u] {} (boundary);
    \draw [->,very thick] (boundary) -- node[name=u] {} (Final_sol);
    \draw [->,very thick] (distance) -- node[name=u] {} (Final_sol);
    \draw [->,very thick] (sol) -- node[name=u] {} (Final_sol);
    
    \draw [->,very thick] (FF) -- node[name=u] {} (sol);
    \draw [->,very thick] (SF) -- node[name=u] {} (sol);
\end{tikzpicture}
    \centering
    
    \caption{The architecture of the PINN used for the electromagnetic interface problem.}
    \label{fig:net_arch2}
\end{figure}
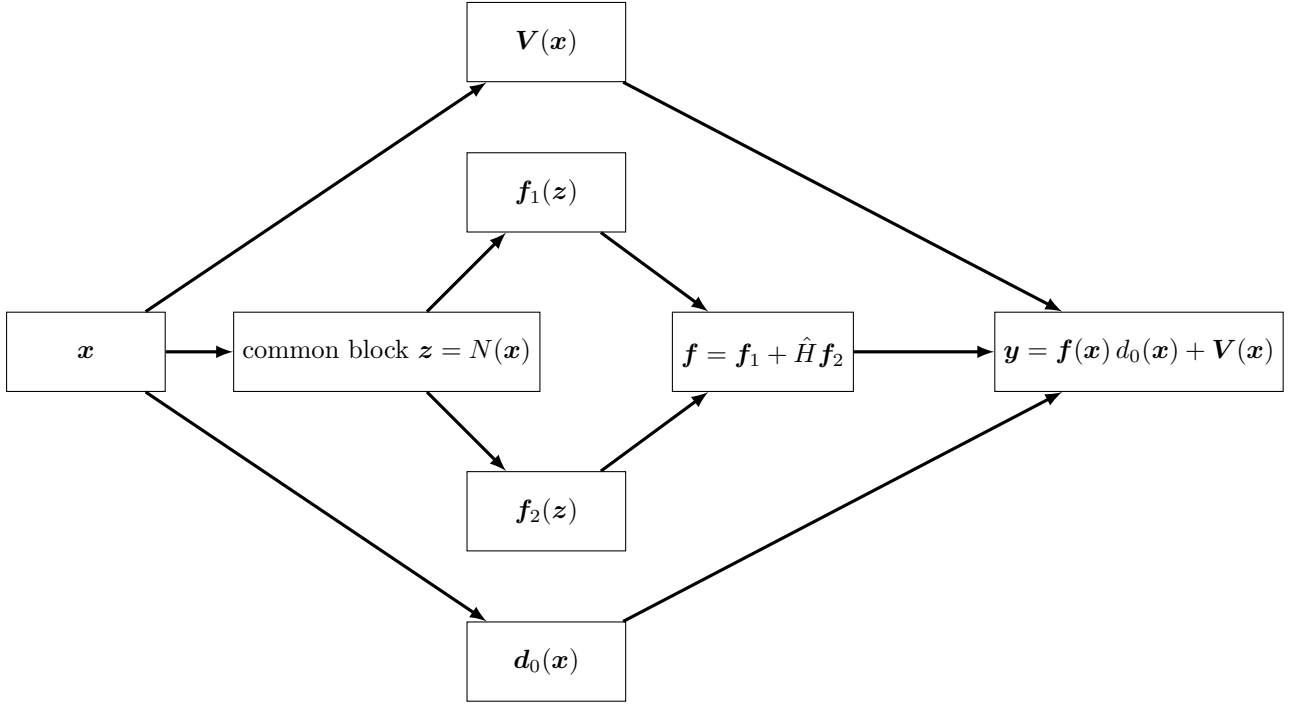

Another problem encountered with PINNs, for interface problems, is that neural networks face a challenge for approximating high-frequency functions \cite{spec_bias,spec_bias2}, such as functions containing sharp gradients. Though much research has been conducted on PINNs to approximate such functions, training these networks is still time-consuming. For electromagnetic problems with material interfaces, the location of the discontinuity, and the jump in the electromagnetic fields are known. Hence, we can exploit these features to build a neural network with improved convergence properties.
To do so, we modify the architecture of the neural network $N(\x)$, illustrated in figure~\ref{fig:net_arch2}, where $\hat{H}$ is the smoothed Heaviside function, $d_0(\x)$ is defined as $d_0 = (1-e^{-10 D(\x)})$ where $D(\x)$ is the distance function to the boundary,  $\boldsymbol{y} = [\boldsymbol{H}, \boldsymbol{E}]$, and $\boldsymbol{V}$ is a lift vector function that satisfies the boundary conditions on $\boldsymbol{H}$ and $\boldsymbol{E}$. 

The PINN will be trained using the loss function defined as\[L= \|\partial_t \mu\boldsymbol{H}+ \nabla \times\boldsymbol{E}  \| + \|- \nabla \times\boldsymbol{H}+ \sigma\boldsymbol{E} \| + \| \nabla \cdot (\mu\boldsymbol{H}) \|  + \|\nabla \cdot (\epsilon\boldsymbol{E}) \| \]
for the transient case, and 
\[L= \| \nabla \times\boldsymbol{E}  \| + \|- \nabla \times\boldsymbol{H}+ \sigma\boldsymbol{E} \| + \| \nabla \cdot (\mu\boldsymbol{H}) \|  + \|\nabla \cdot (\epsilon\boldsymbol{E}) \| \]
for the steady-state case.

\section{Numerical results}

\subsection{Magnetostatics in continuous media}
\begin{figure}
\centering

\begin{tikzpicture}
%cylinder
\draw [thick](-4,-1) -- (-4,1);
\draw [thick](+4,-1) -- (+4,1);
\draw [thick](-4,-1) arc (180:360:4 and 1);          % <--
\draw[thick,dashed] (4,-1) arc (-1:180:4 and 1);  % <--
\draw [thick](-4,+1) arc (180:360:4 and 1);          % <--
\draw [thick](+4,+1) arc (-4:180:4 and 1);         % <--
\draw[thick,gray,dashed](0,-1) --(+4,-1);
%hollow
\draw [thick](-1,-1) -- (-1,1);
\draw [thick](+1,-1) -- (+1,1);
\draw [thick](-1,-1) arc (180:360:1 and 0.25);
\draw[thick,dashed] (1,-1) arc (-1:180:1 and 0.25);
\draw [thick](-1,+1) arc (180:360:1 and 0.25);
\draw [thick](+1,+1) arc (-1:180:1 and 0.25);
\draw[thick,gray,dashed](0,-1) --(+1,-1);
%
%\fill[fill=black] (0,-1.65) circle (1.5pt);
%\node[below,scale=0.8] at (1.2,-2) {$R$};
%%
%% I changed a little the node position; removed \text{}
%% node for R_1 below
\draw [thick,<->] (0,-1) -- node[fill=white,inner sep=1pt] {\scriptsize$0.25$} (1,-1);
\draw [thick,<->] (0,-2.5) -- node[fill=white,scale=0.9,inner sep=1pt] {$1$ } (4,-2.5);
\draw [thick,<->] (4.5,1) -- node[fill=white,scale=0.9, inner sep=1pt] {$0.25$ } (4.5,-1);
\end{tikzpicture}

\caption{The geometry for the Ampère problem (length units are nondimensional) .}
\label{fig:cyl_geo}

\end{figure}
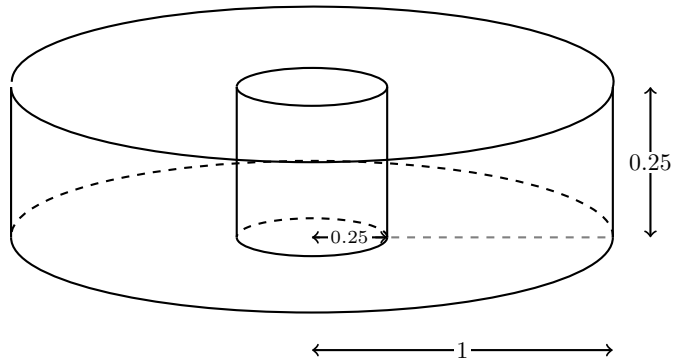

\begin{figure}[H]
\begin{subfigure}[t]{0.5\textwidth}
\centering
    \includegraphics[width=0.9\textwidth,height=0.9\textwidth]{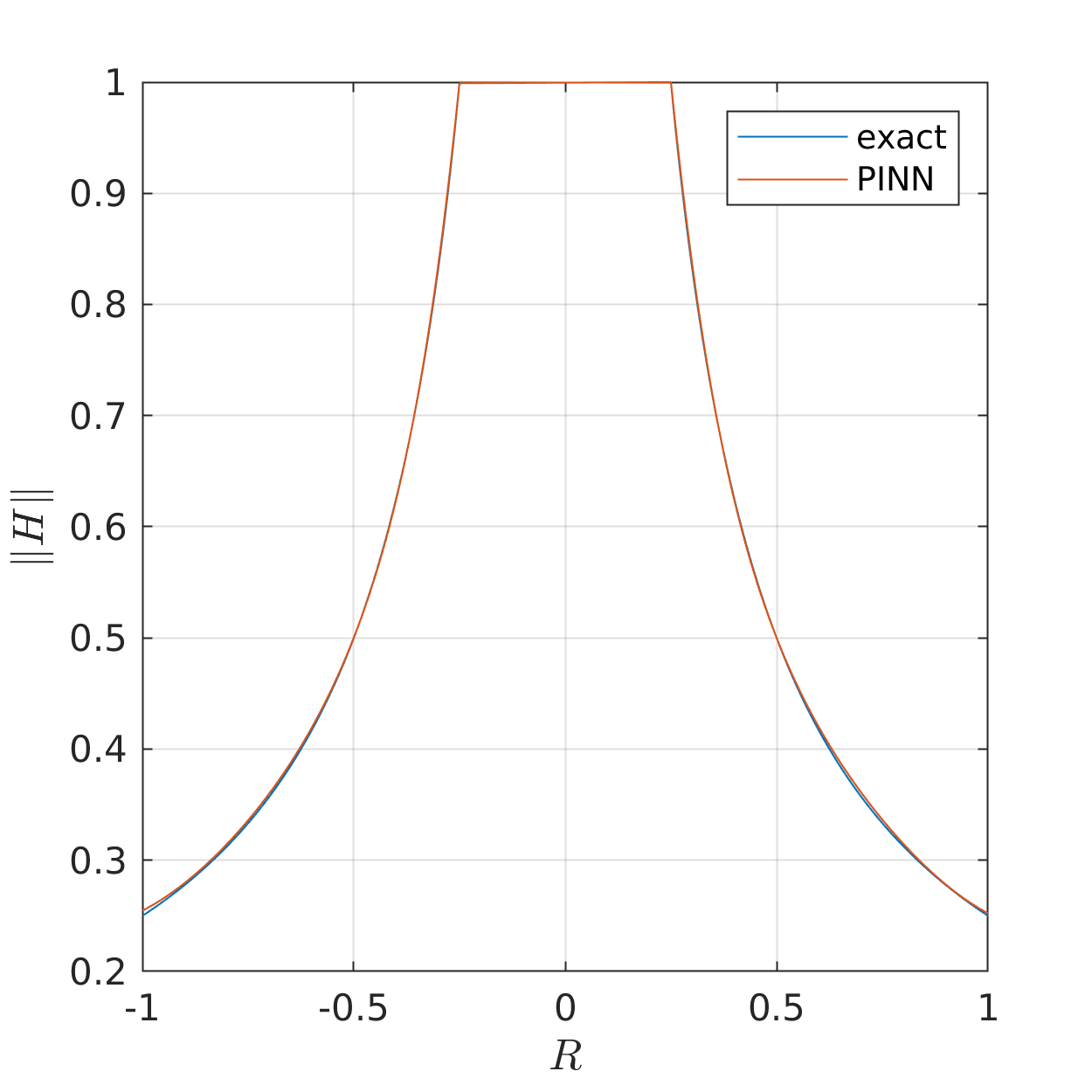}

    \caption{}
\end{subfigure}
\begin{subfigure}[t]{0.5\textwidth}
\centering
    \includegraphics[width=0.9\textwidth,height=0.9\textwidth]{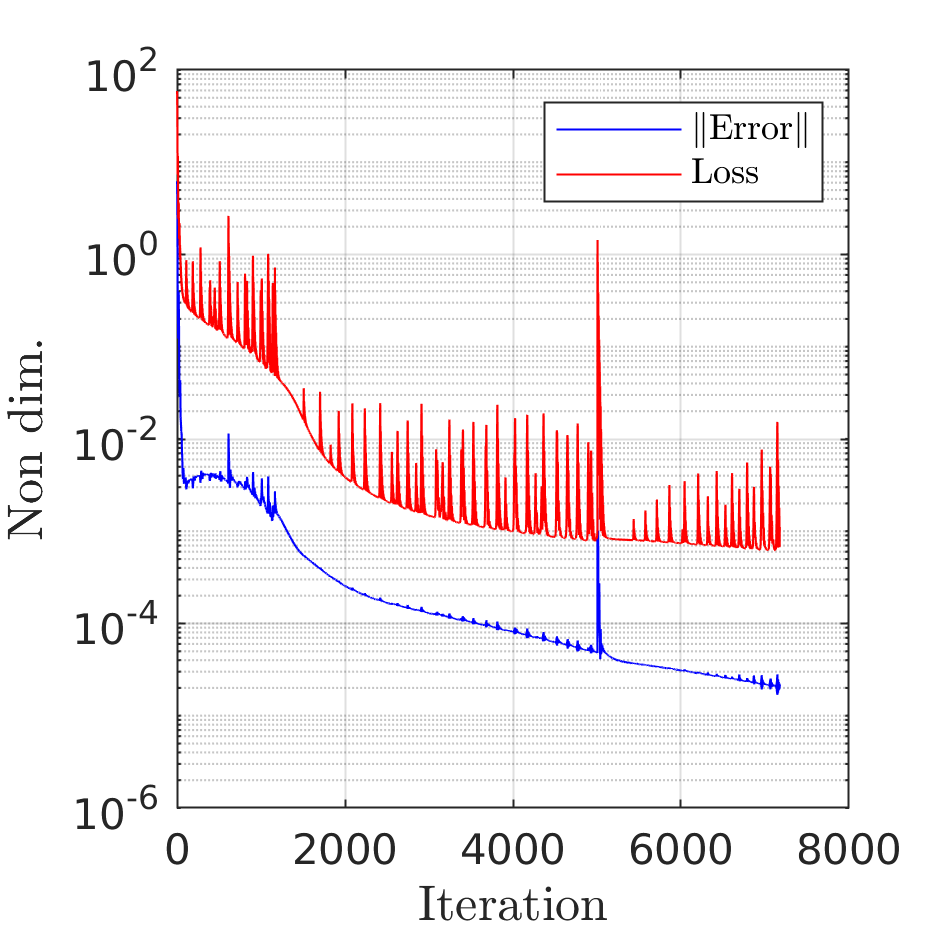}

    \caption{}
\end{subfigure}

    \caption{The results obtained for the steady-state Ampère problem. (a) The magnitude $\|\boldsymbol{H}\| $ compared to the analytical solution (b) The norm of the loss function and the error during training. }
    \label{fig:st-con}
  \end{figure}

Our study begins with the study of the steady-state Ampère problem in a continuous media. The geometry depicted in figure~\ref{fig:cyl_geo} consists in a hollow cylinder with an inner radius of 0.25 nondimensional units, and an outer radius of 1.
On the cylinder's inner surface, we enforce a nondimensional Dirichlet boundary condition given by \mbox{$\boldsymbol{H}=4 \, (-y, x,0)$}. For all other surfaces, the Dirichlet boundary condition $\boldsymbol{E}=(0,0,0)$ is applied. Figure~\ref{fig:st-con} shows how the error on the solution decreases with the number of iterations, the correlation between the norm of the loss function during training, and the norm of the error. The final results obtained after 7000 Adam training iterations are shown in figure~\ref{fig:st-con}, and have a maximum relative error of $0.4 \% .$ 

\subsection{Steady-state magnetic problem with discontinuous media}

\subsubsection{2D steady-state problem}
\begin{figure}[H]
\begin{subfigure}[t]{0.32\textwidth}
\centering
    \includegraphics[width=\textwidth, height=\textwidth]{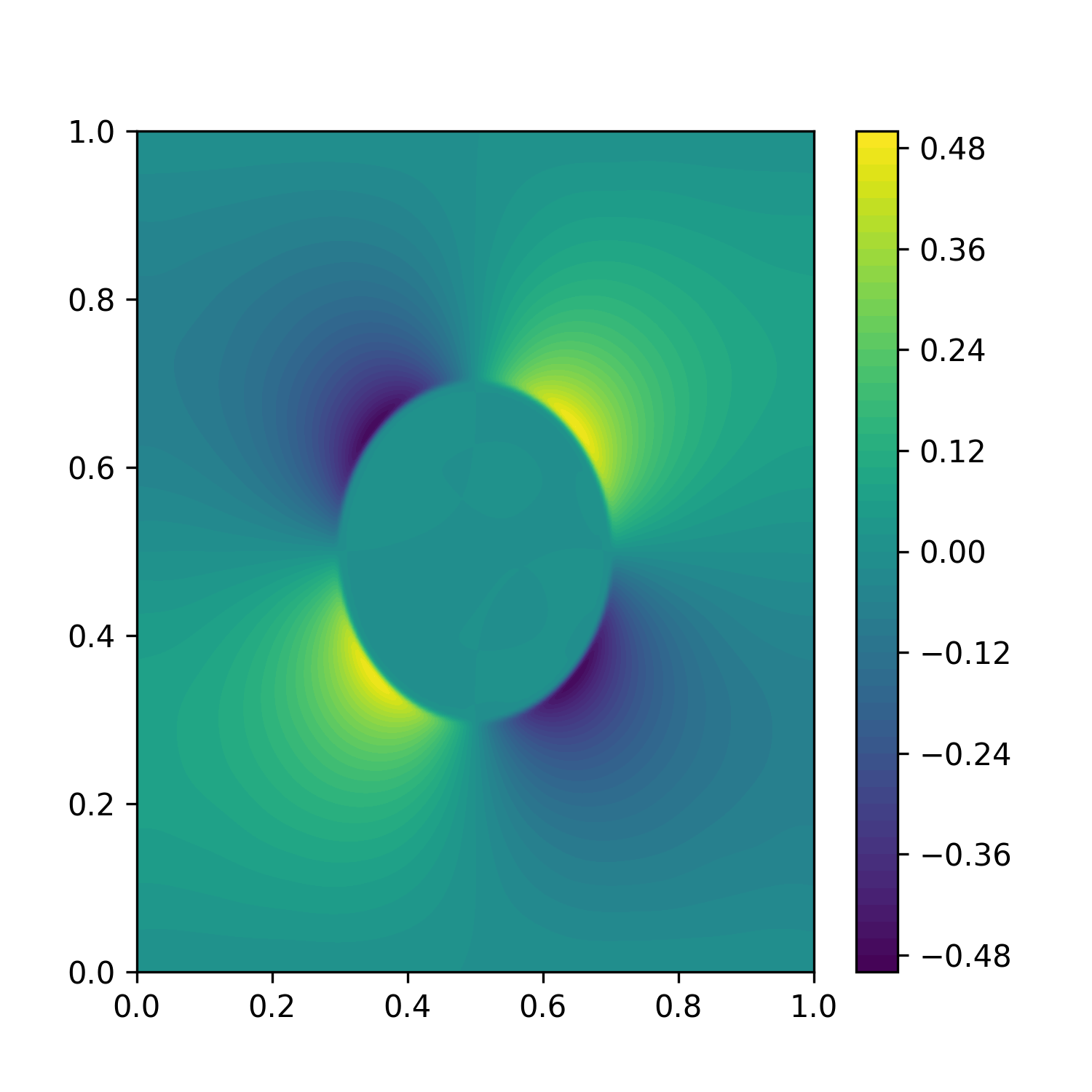}
    \caption{}
\end{subfigure}
\begin{subfigure}[t]{0.32\textwidth}
    \includegraphics[width=\textwidth , height=\textwidth ]
    {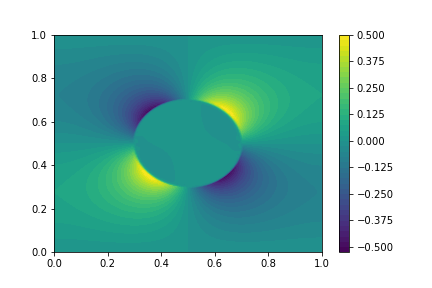}
    \caption{}
\end{subfigure}
\begin{subfigure}[t]{0.32\textwidth}
\centering
    \includegraphics[width=\textwidth, height=\textwidth]{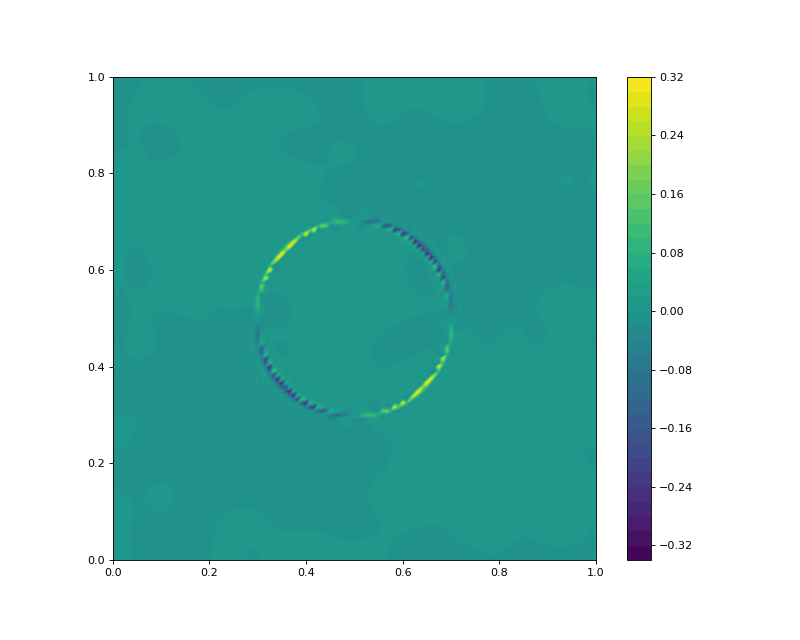}
    \caption{}
\end{subfigure}

    \caption{The $x$-component of the magnetic field $H_x$ for the 2D steady-state interface problem of a disc inside a unit square (a) using the PINN method (b) using the FEM (c) the difference between the two results.}
    \label{r1}

\begin{subfigure}[t]{0.32\textwidth}
\centering
    \includegraphics[width=\textwidth, height=\textwidth]{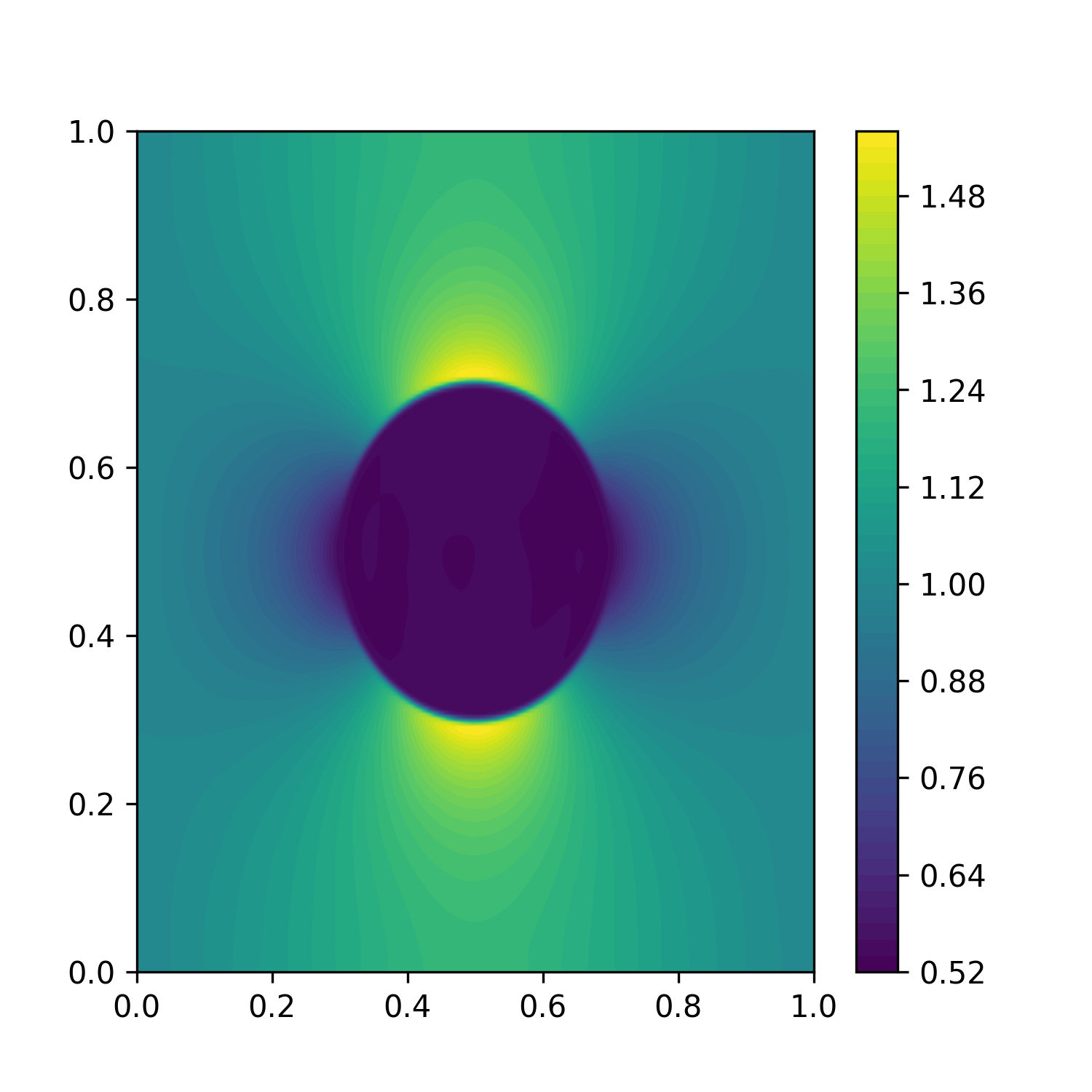}
    \caption{}
\end{subfigure}
\begin{subfigure}[t]{0.32\textwidth}
    \includegraphics[width=\textwidth, height=\textwidth]{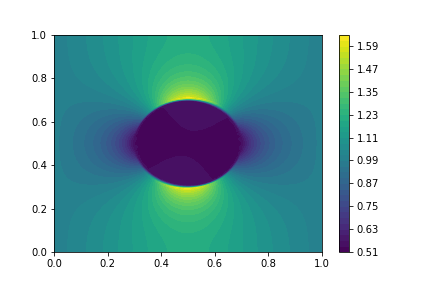}
    \caption{}
\end{subfigure}
\begin{subfigure}[t]{0.32\textwidth}
    \includegraphics[width=\textwidth, height=\textwidth]{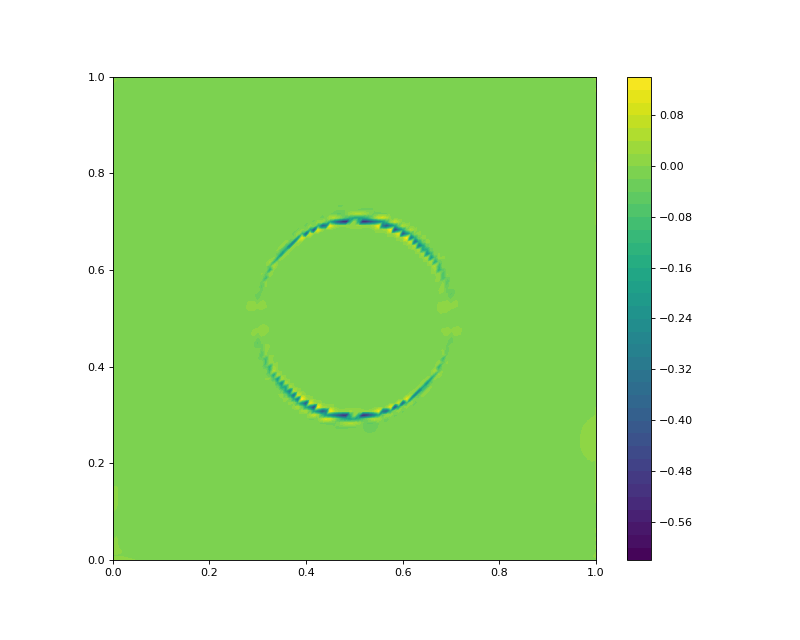}
    \caption{}
\end{subfigure}
    \caption{The $z$-component of the magnetic field $H_z$ for the 2D steady-state interface problem of a disc inside a unit square (a) using the PINN method (b) using the FEM (c) the difference between the two results.}
    \label{r2}

\end{figure}

We now verify the proposed PINN architecture on the 2D steady-state problem consisting of a disc with a radius of 0.2 nondimensional unit inside a unit square. The Dirichlet boundary condition $\boldsymbol{H}=[0,0,1]$ is enforced on the left and the right sides of the square. The permeability of the sphere is 3 nondimensional units, and the permeability outside the sphere is 1 nondimensional unit. The Heaviside function is approximated using a sigmoid function of the form $\mu=S ((0.2-r) 500 )$. We compare, in figures \ref{r1} and \ref{r2}, the results obtained using the proposed PINN trained with 3000 Adam iterations, with the results obtained using the finite element method.

As we can see for the two-dimensional test case, the PINN converges toward the approximation obtained using the FEM, but with a steeper gradient at the interface. This is due to the meshless and continuous representation computed by the PINN, as opposed to the FEM, where the elements size add a constraint on the interface representation.

\subsubsection{3D steady-state and parametric problem}
\begin{figure}
\begin{subfigure}[t]{0.5\textwidth}
\centering
    \includegraphics[width=\textwidth,height=\textwidth]{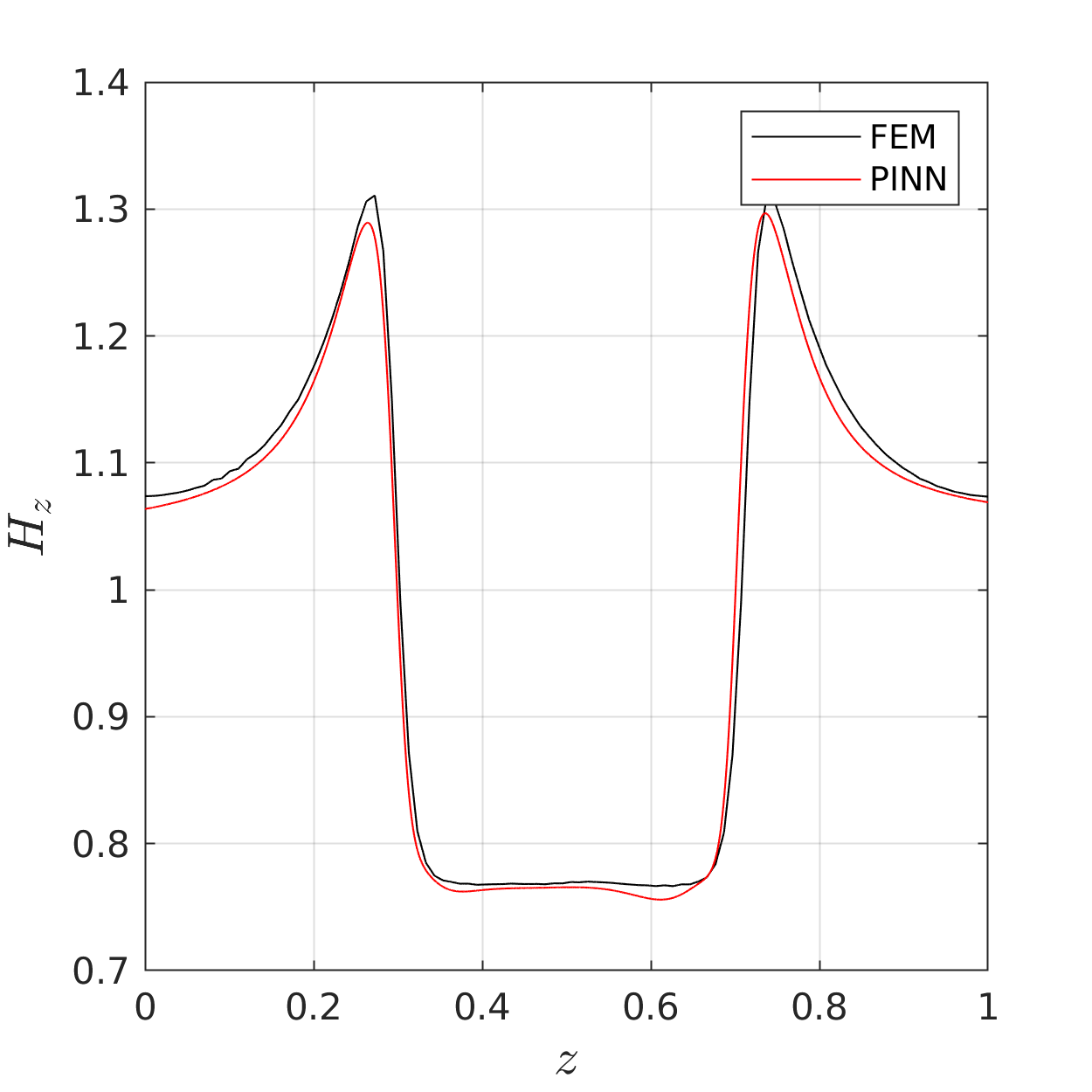}
    \caption{}
\end{subfigure}
\begin{subfigure}[t]{0.5\textwidth}
    \includegraphics[width=\textwidth,height=\textwidth]{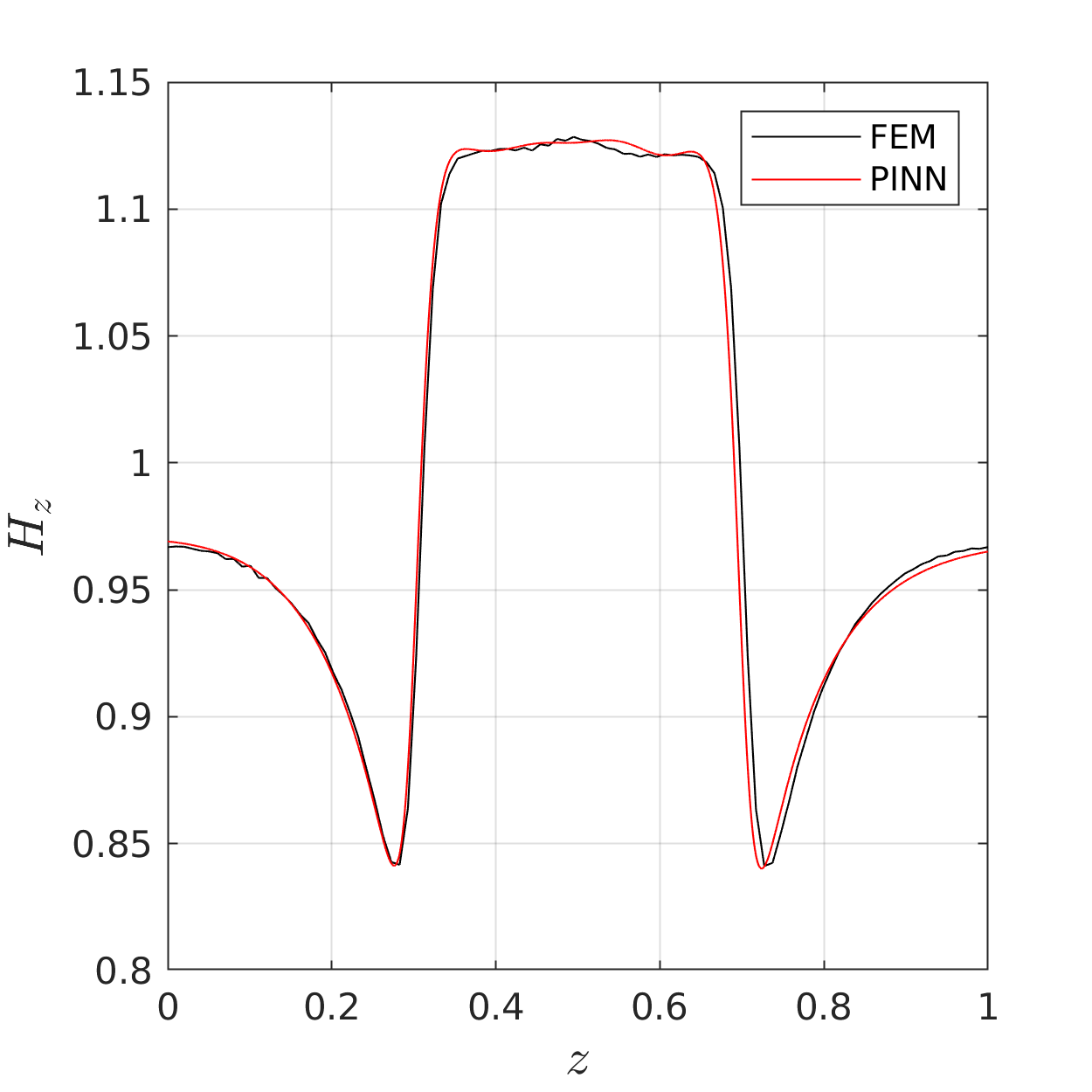}
    \caption{}
\end{subfigure}
\begin{subfigure}[t]{\textwidth}
    \centering  \includegraphics[width=0.5\textwidth,height=0.5\textwidth]{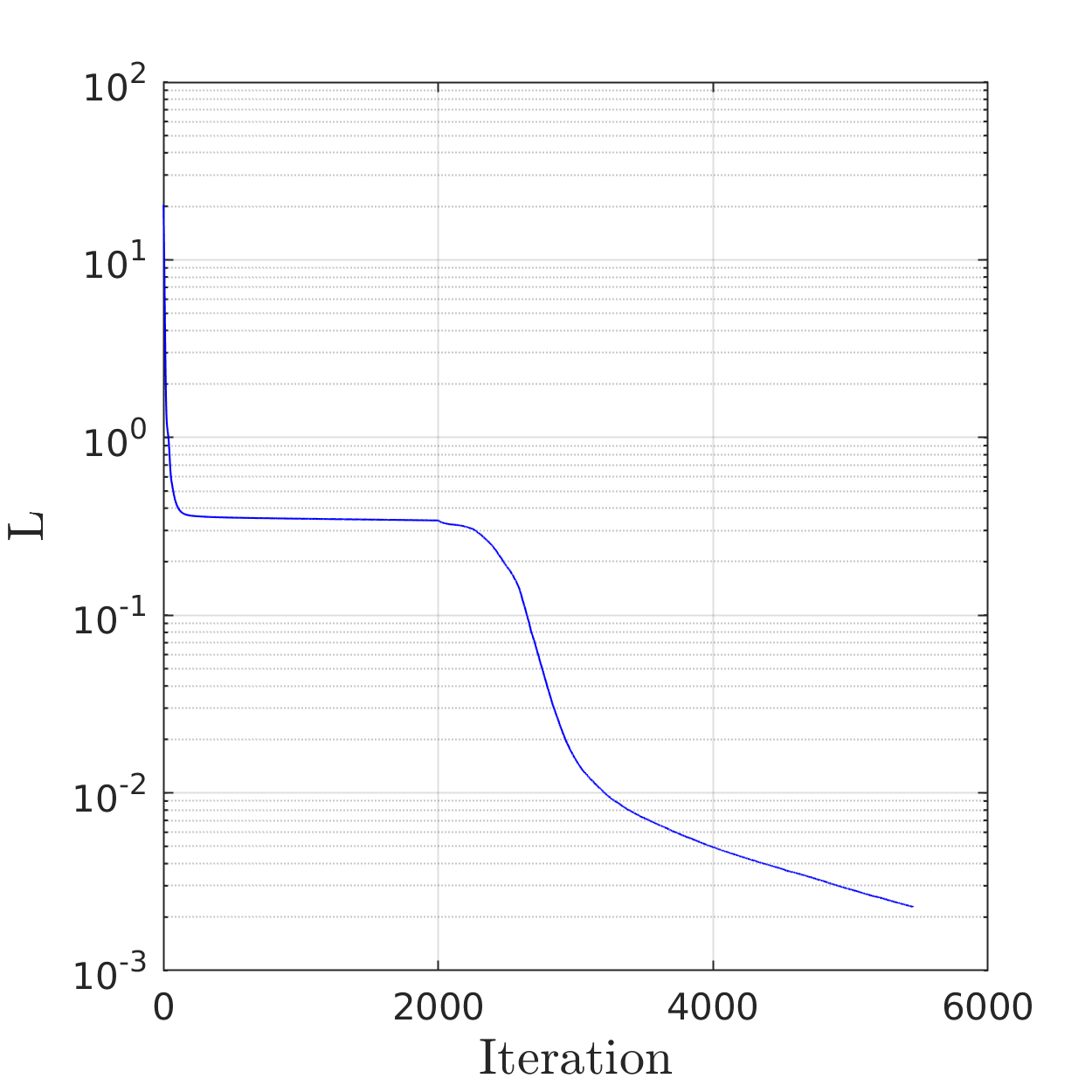}
    \caption{}
\end{subfigure}
    \caption{The $z$-component of the magnetic field $H_z$ obtained using the proposed PINN method and the FEM, for the steady-state parametric problem of a sphere inside a unit cube: (a) $\mu_{\rm out} =0.5$; (b) $\mu_{\rm out} =1.5$; (c) Evolution of the loss function during training with 2000 Adam iterations and 3500 L-BFGS iterations.}
    \label{fig:disc_mu}
  \end{figure}

In this numerical test, we apply the proposed PINN methodology to the parametric 3D problem of a sphere inside a unit cube. The inputs of the PINN are the spatial coordinates and the permeability value outside the sphere, which ranges from $0.5$ to $1.5$ nondimensional units. The sphere's permeability is constant and equal to $1$. We approximate the Heaviside function using $\hat{H}= S( 100(0.2 - r))$. The results, shown in figure~\ref{fig:disc_mu}, obtained using the proposed PINN, show good agreement with the results obtained with the FEM.

\begin{figure}
\begin{subfigure}[t]{0.5\textwidth}
\centering
    \includegraphics[width=\textwidth]{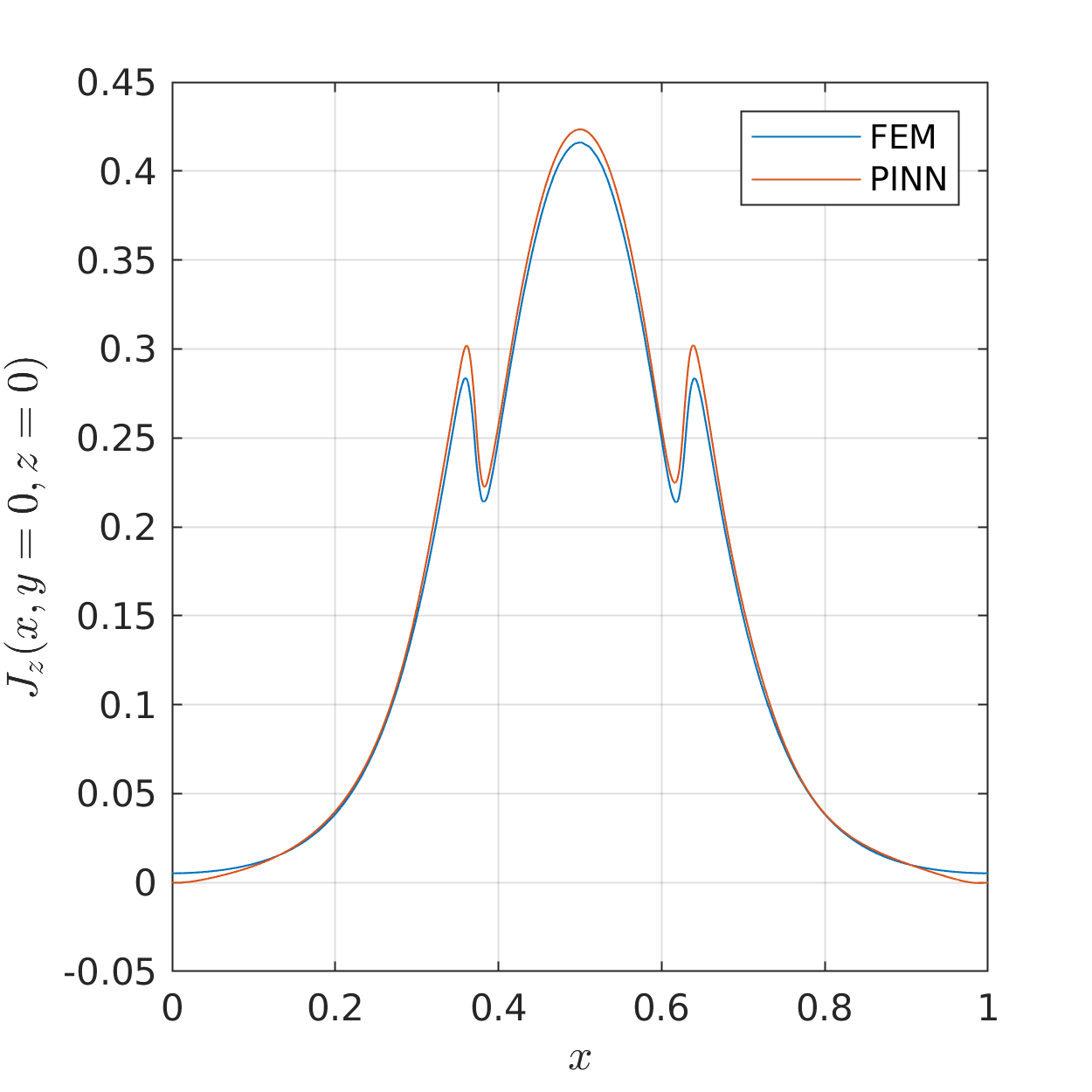}
\caption{}
\end{subfigure}
\begin{subfigure}[t]{0.5\textwidth}
    \includegraphics[width=\textwidth]{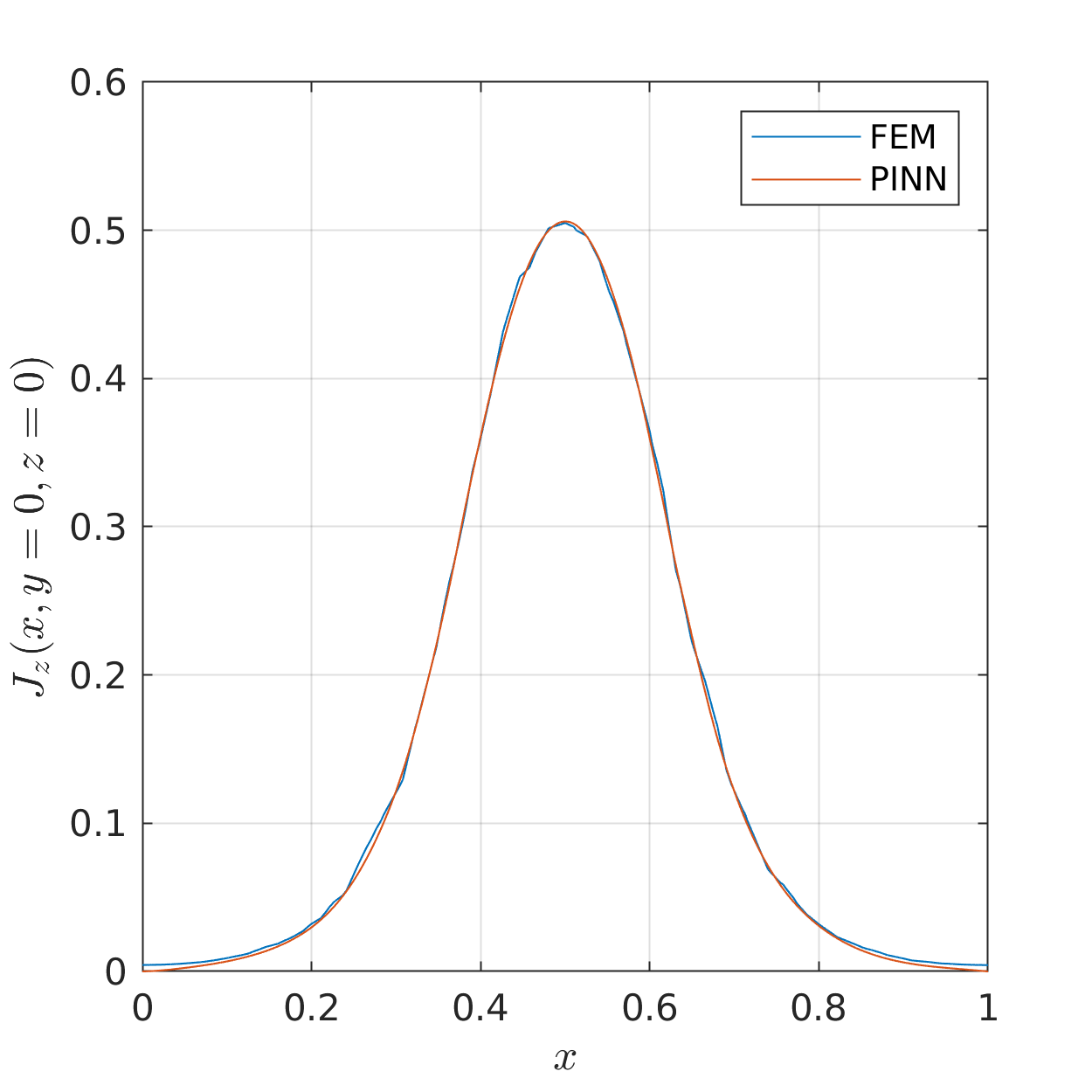}
\caption{}
\end{subfigure}
\begin{subfigure}[t]{0.5\textwidth}
    \includegraphics[width=\textwidth]{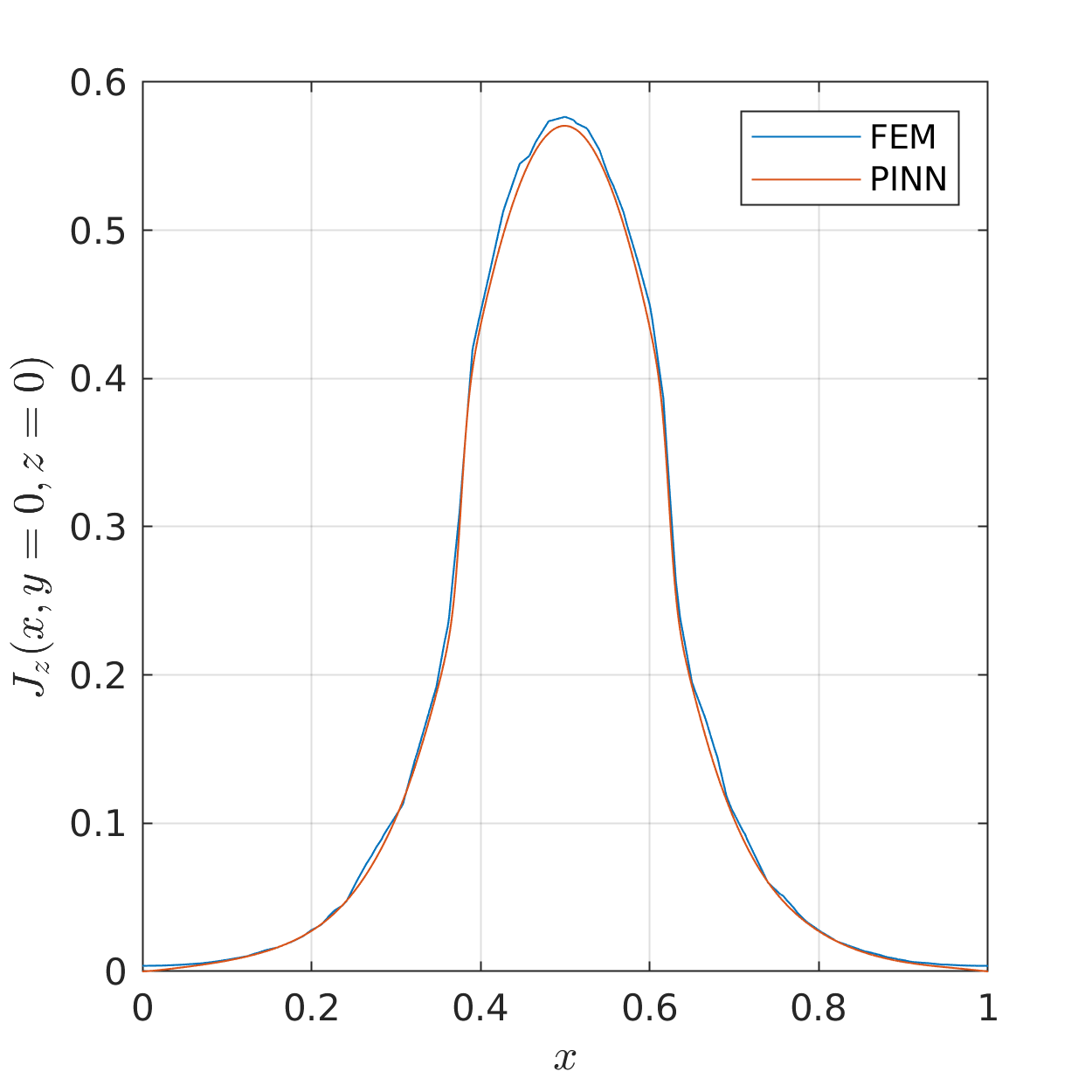}
    \caption{}
\end{subfigure}
\begin{subfigure}[t]{0.5\textwidth}
    \includegraphics[width=\textwidth]{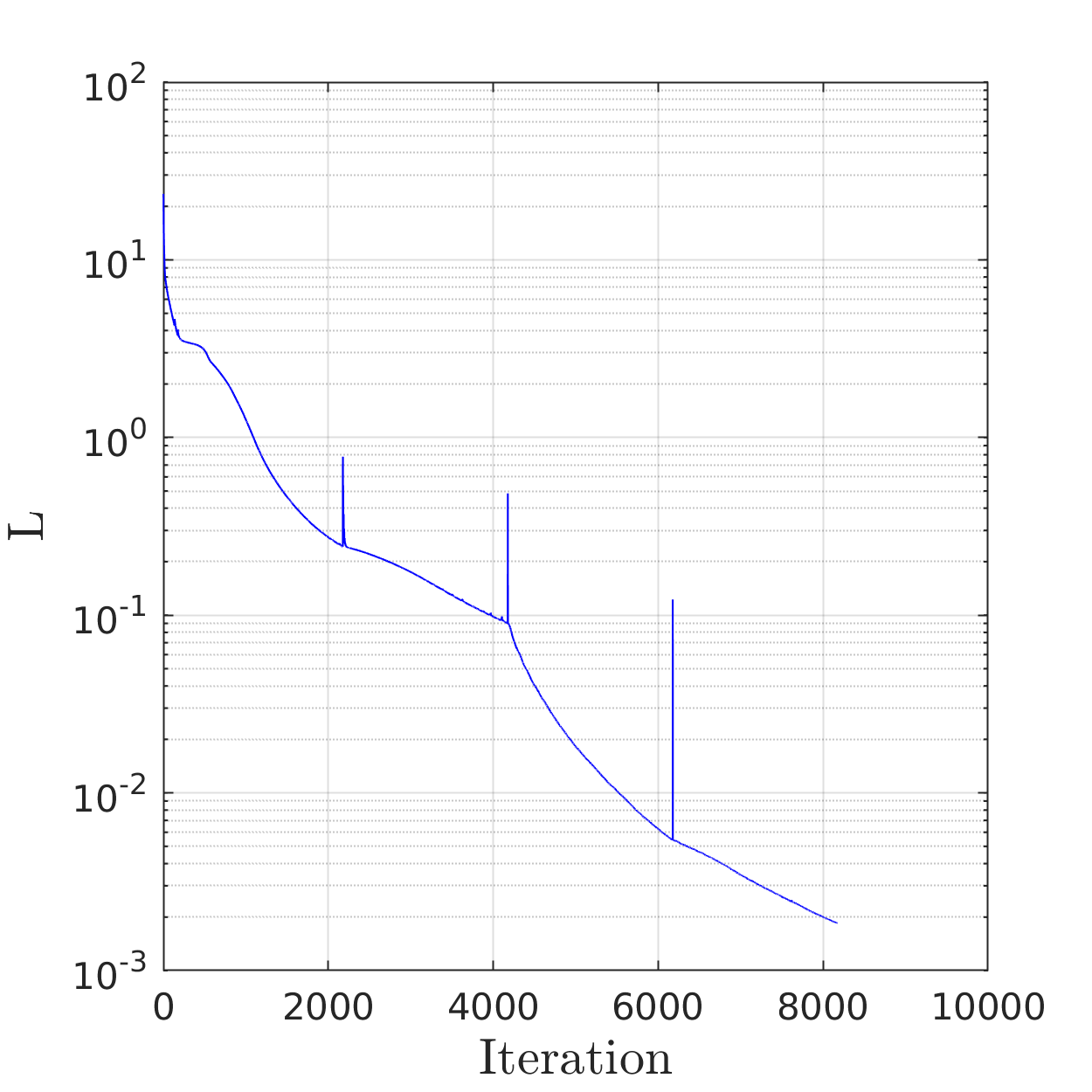}
    \caption{}
\end{subfigure}
    \caption{The $z$-component of the electric current $J_z$ at $y=z=0$, for the steady-state parametric problem with discontinuous resistivity: (a) $\rho_{\rm out}=0.5 $; (b) $\rho_{\rm out}=1$; (c) $\rho_{\rm out}=1.5$; (d) Loss function evolution during training with 4000 Adam iterations and 4000 L-BFGS iterations.}
    \label{fig:disc_rho}
  \end{figure}

In our next experiment, we test the steady-state electromagnetic solver with a discontinuous resistivity. The geometry consists in two concentric cylinders, with radii 0.25 and 1 nondimensional unit. On the bottom of both cylinders, we impose the Dirichlet boundary conditions $E_x=0$ and $E_y=0$. On the top of the inner cylinder, we impose $E_z=1$ and on the rest of the boundaries, we impose $E_z=0$.
The inputs of the PINN are the spatial coordinates and the resistivity value of the outer cylinder, which ranges from $0.5$ to $1.5$ nondimensional units. The inner cylinder's resistivity is constant and equal to $1$.
We compare the results obtained using the proposed PINN with the results obtained using the FEM in figure~\ref{fig:disc_rho}, and we see good agreements between the results. Additionally, the results obtained using the PINN verify the boundary conditions, while the results  obtained using the FEM, where the boundary conditions on $\mathbf{E}$ are treated as Neuman boundary conditions on the scalar potential $V$, do not exactly satisfy the boundary conditions.

\subsection{Transient and parametric electromagnetic problems with discontinuous media}

\begin{figure}[ht]
\begin{subfigure}[t]{0.5\textwidth}
\centering
    \includegraphics[width=7cm,height=7cm]{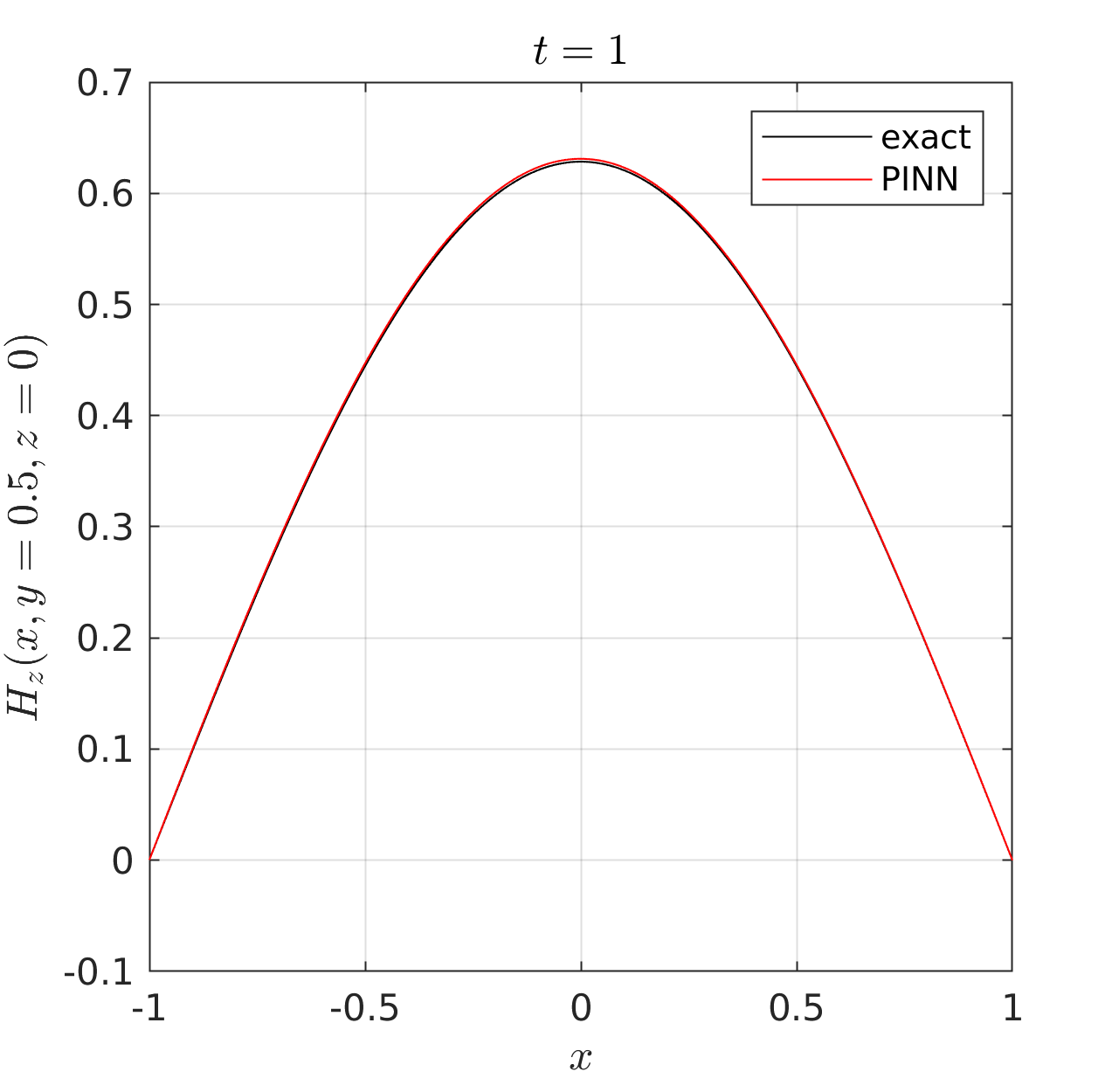}
    \caption{}
\end{subfigure}
\begin{subfigure}[t]{0.5\textwidth}
\centering
    \includegraphics[width=7cm,height=7cm]{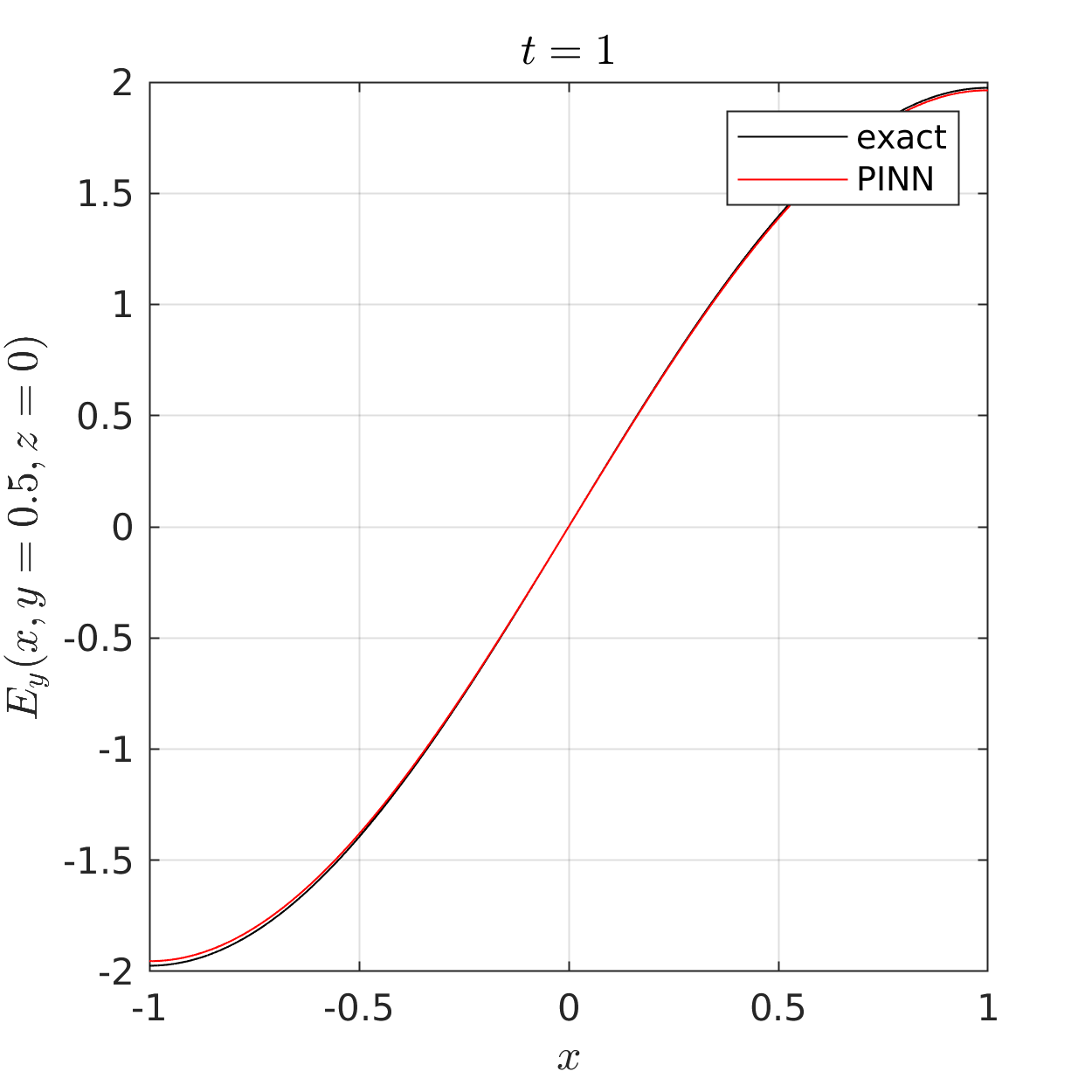}
    \caption{}
\end{subfigure}
    \caption{  (a) The $z$-component of the magnetic field ${H}_z$; (b) the $y$-component of the electric field ${E}_y$ for the first transient problem with a manufactured solution using a PINN at $t=1$, $y=0.5$ and $z=0$. }
    \label{fig:manu2}
  \end{figure}

In this section, we test the proposed PINN on transient problems.
First, we consider a problem with a manufactured solution with constant permeability and resistivity, and no material interface. The geometry is a unit cube. The solution is given by $$ \boldsymbol{H}=[0,0, \sin(\pi x) \sin(\pi y)]  (1-e^{-t})$$
and
$$\boldsymbol{E}= [  \pi \sin(\pi x) \cos(\pi y) ,-\pi \cos(\pi x) \sin(\pi y),0]  (1-e^{-t}) .$$
Figure~\ref{fig:manu2} shows a good agreement between the results obtained using the PINN and the exact solution, with a maximum relative error of $2 \times 10^{-3}.$

\begin{figure}[H]
\begin{subfigure}[t]{0.5\textwidth}
\centering
    \includegraphics[width=7cm,height=7cm]{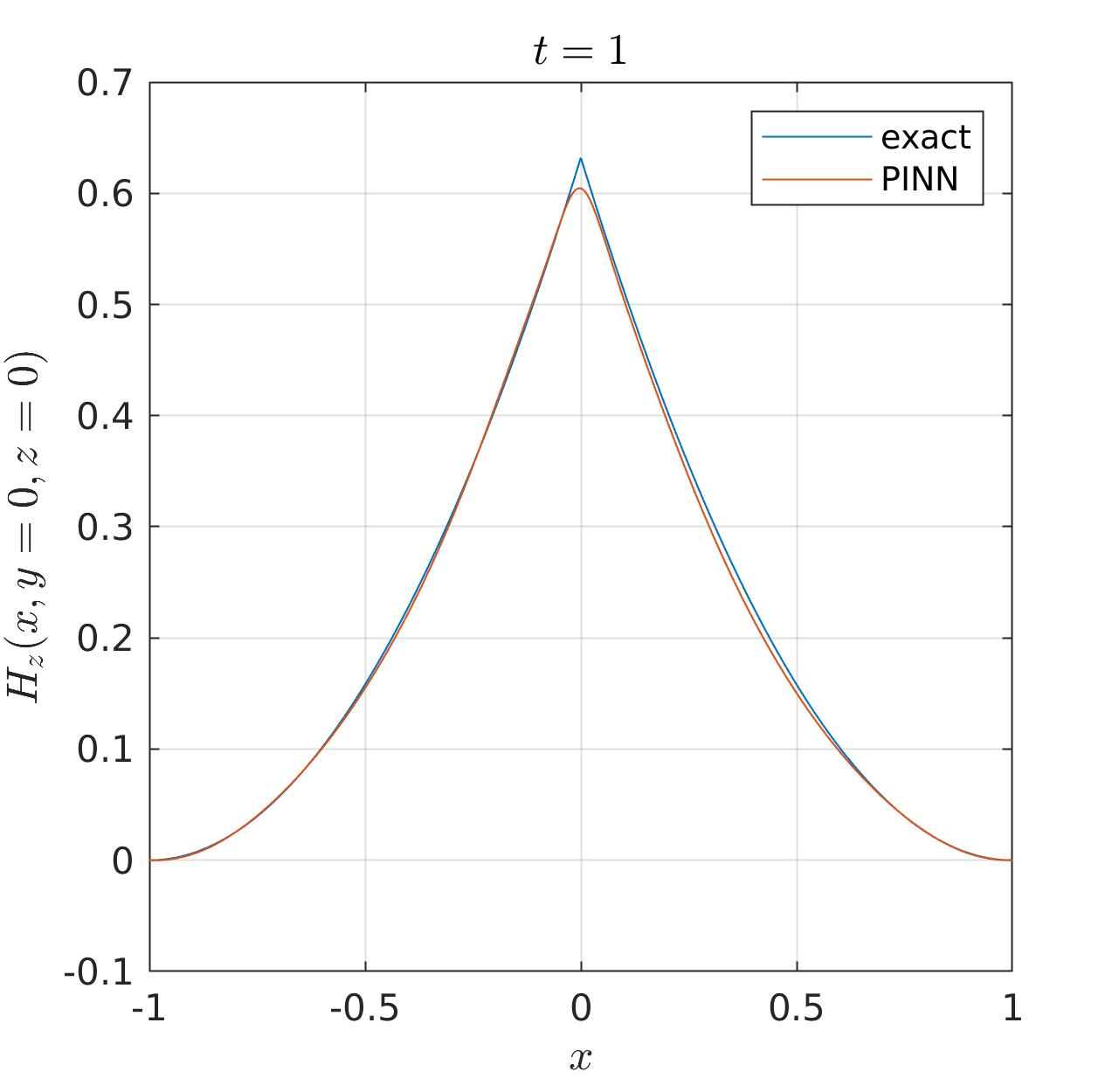}
    \caption{}
\end{subfigure}
\begin{subfigure}[t]{0.5\textwidth}
\centering
    \includegraphics[width=7cm,height=7cm]{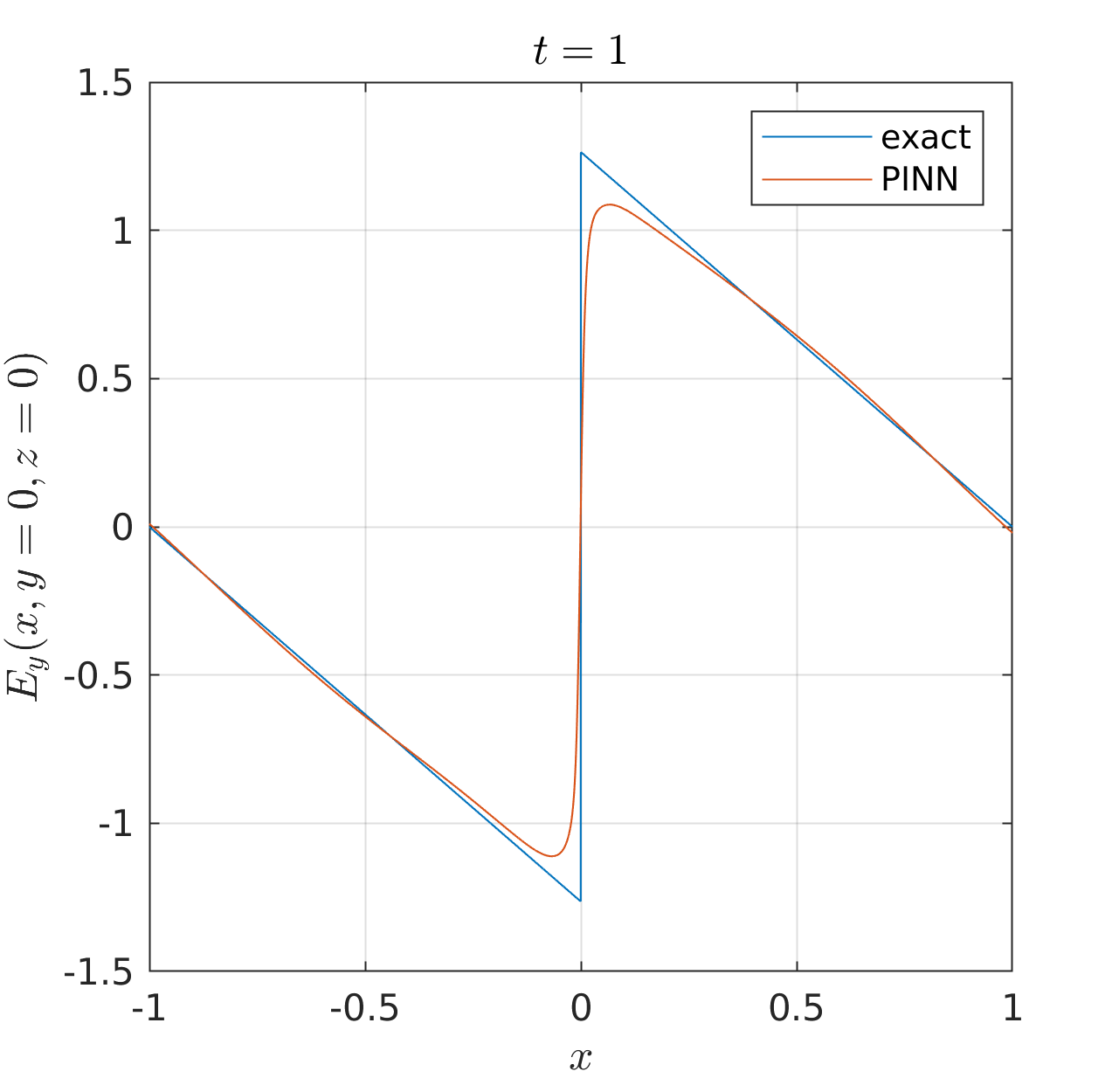}
    \caption{}
\end{subfigure}
    \caption{ Comparison between the analytic solution and the results obtained using the PINN for the second transient problem with a manufactured solution at $t=1$, $y=0$ and $z=0$: (a) $z$-component of the magnetic field ${H}_z$; (b) $y$-component of the electric field ${E}_y$. }
    \label{fig:manu1}
  \end{figure}

Next, we consider a second problem with a manufactured solution given by $$\boldsymbol{H}=[0,0,(1- \sqrt{x^2 + y^2})^2 ] (1-e^{-t})$$
and
$$\boldsymbol{E}=[(2-2/ \sqrt{x^2 + y^2} ) y , (-1+ 1/ \sqrt{x^2 + y^2} ) x , 0 ]  (1-e^{-t}) .$$
The geometry is a cylinder with radius and height $r=h=1$ nondimensional units.
The boundary condition is $\boldsymbol{H}=\boldsymbol{0}$ for $r=1$, and $E_z=0$ on the top and bottom of the cylinder, and with initial condition $\boldsymbol{H}=\boldsymbol{0}$.
The analytic solution contains a discontinuity in the electric field, making it an interesting problem to study. The results illustrated in figure~\ref{fig:manu1} show good agreement with the analytic solution, with an error caused by the continuous representation of a discontinuous field.

% We chose this particular solution to avoid introducing information on the solution in the functions $d(x)$ and $V(x)$. The functions are defined \\ as $d(x)=Tanh(50 \cdot(1-r))$ and $V(x)=0$.

\begin{figure}[H]
\begin{subfigure}[t]{0.5\textwidth}
\centering
    \includegraphics[width=7cm,height=7cm]{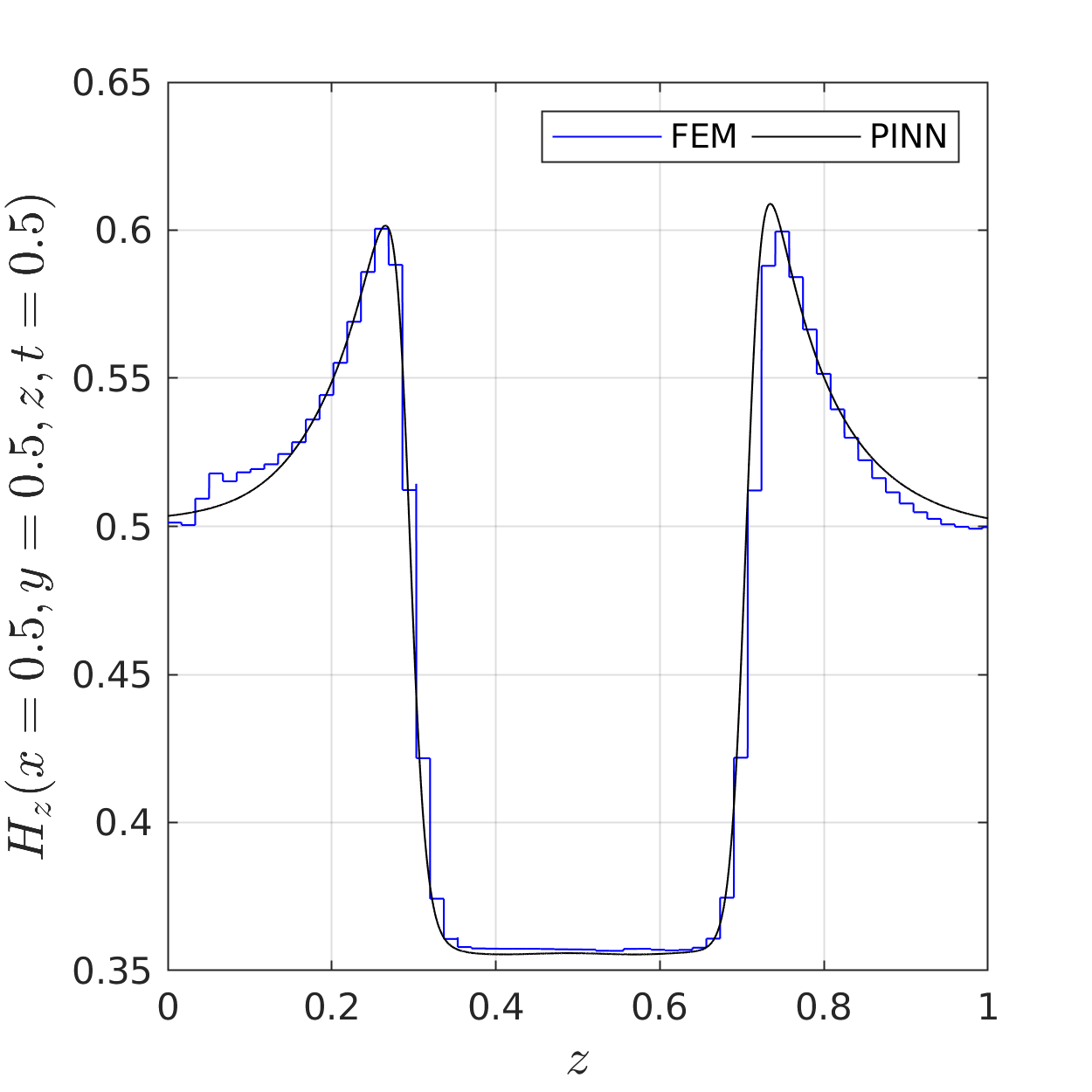}
    \caption{}
\end{subfigure}
\begin{subfigure}[t]{0.5\textwidth}
\centering
    \includegraphics[width=7cm,height=7cm]{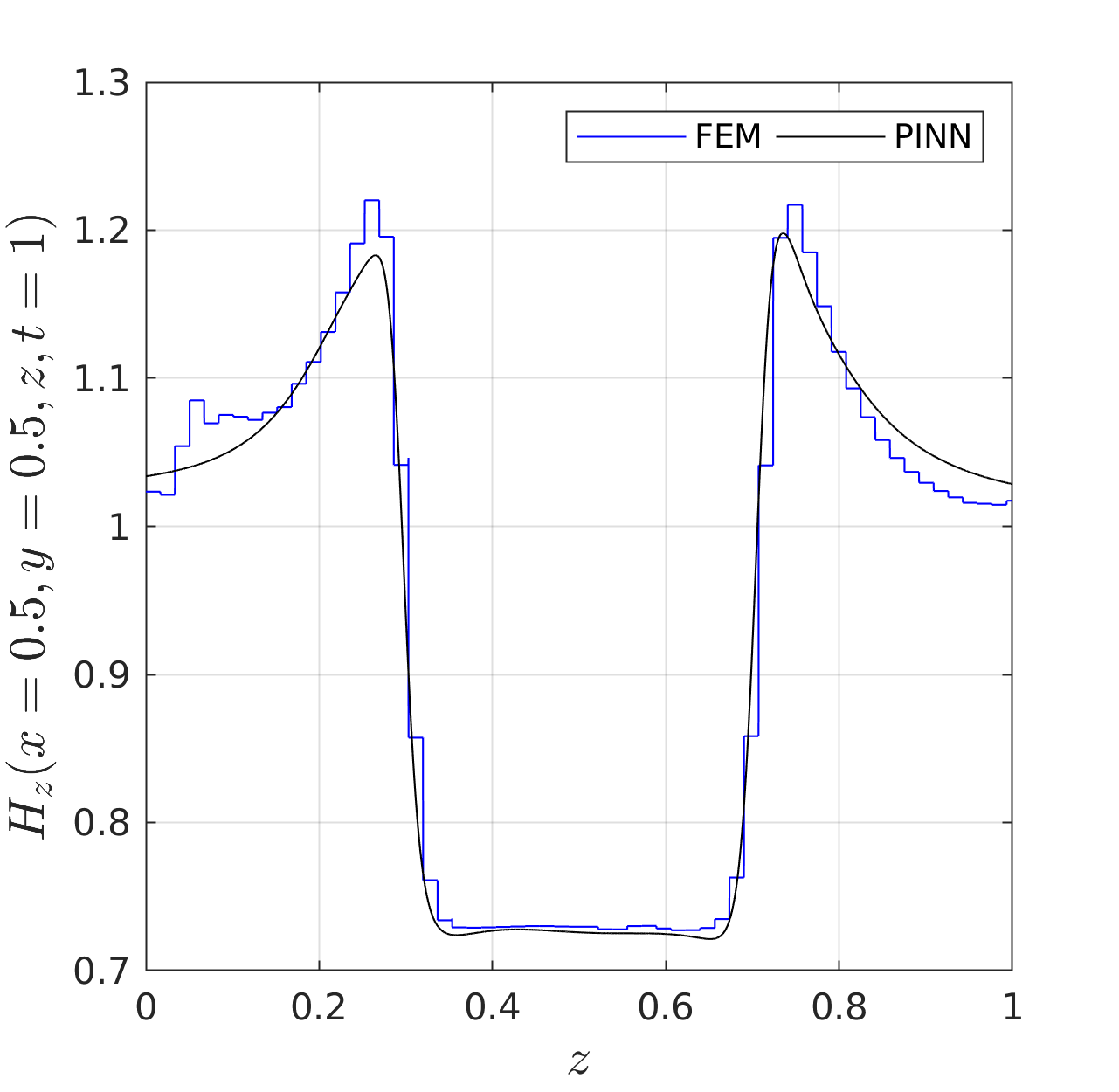}
    \caption{}
\end{subfigure}

\begin{subfigure}[t]{0.5\textwidth}
\centering
    \includegraphics[width=7cm,height=7cm]{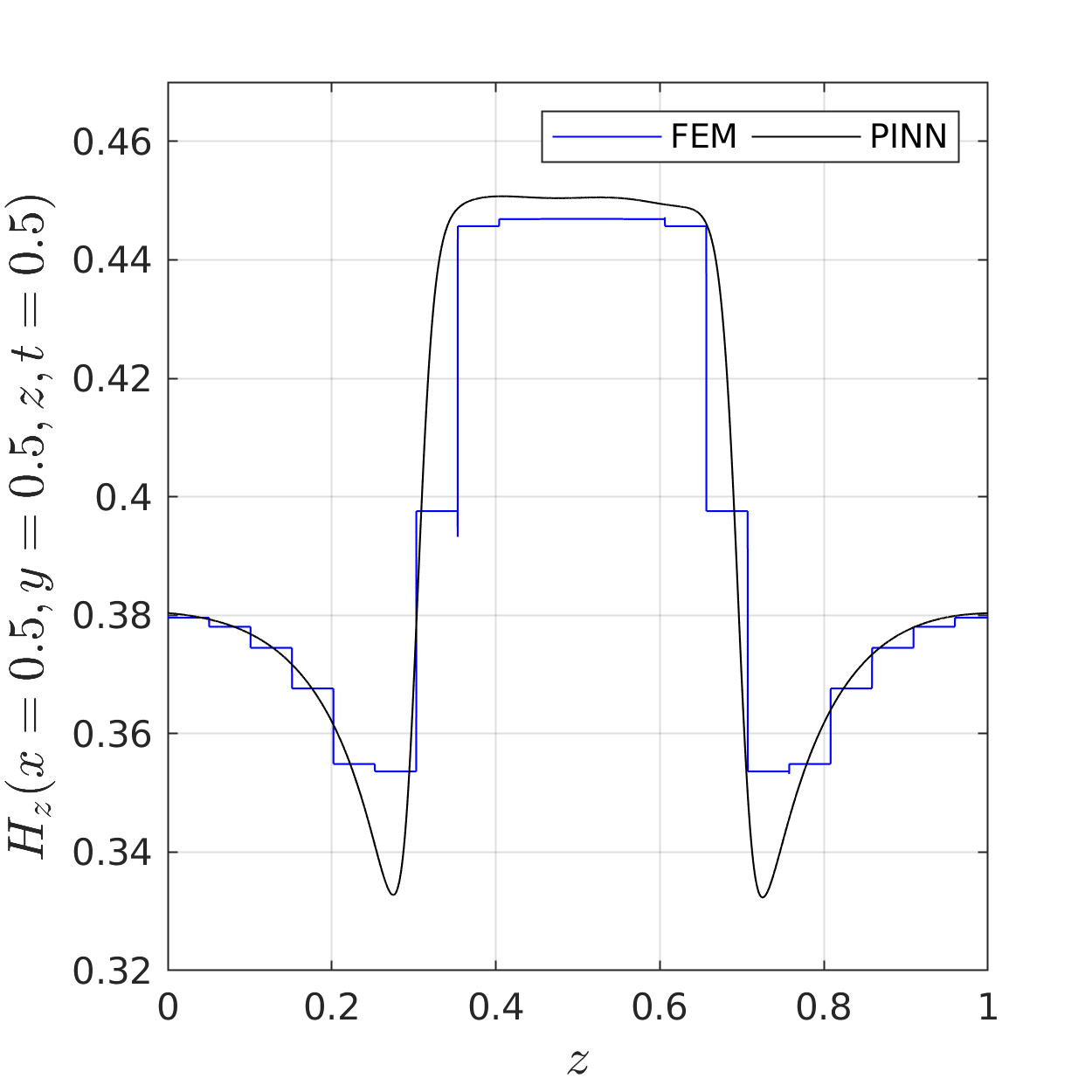}
    \caption{}
\end{subfigure}
\begin{subfigure}[t]{0.5\textwidth}
\centering
    \includegraphics[width=7cm,height=7cm]{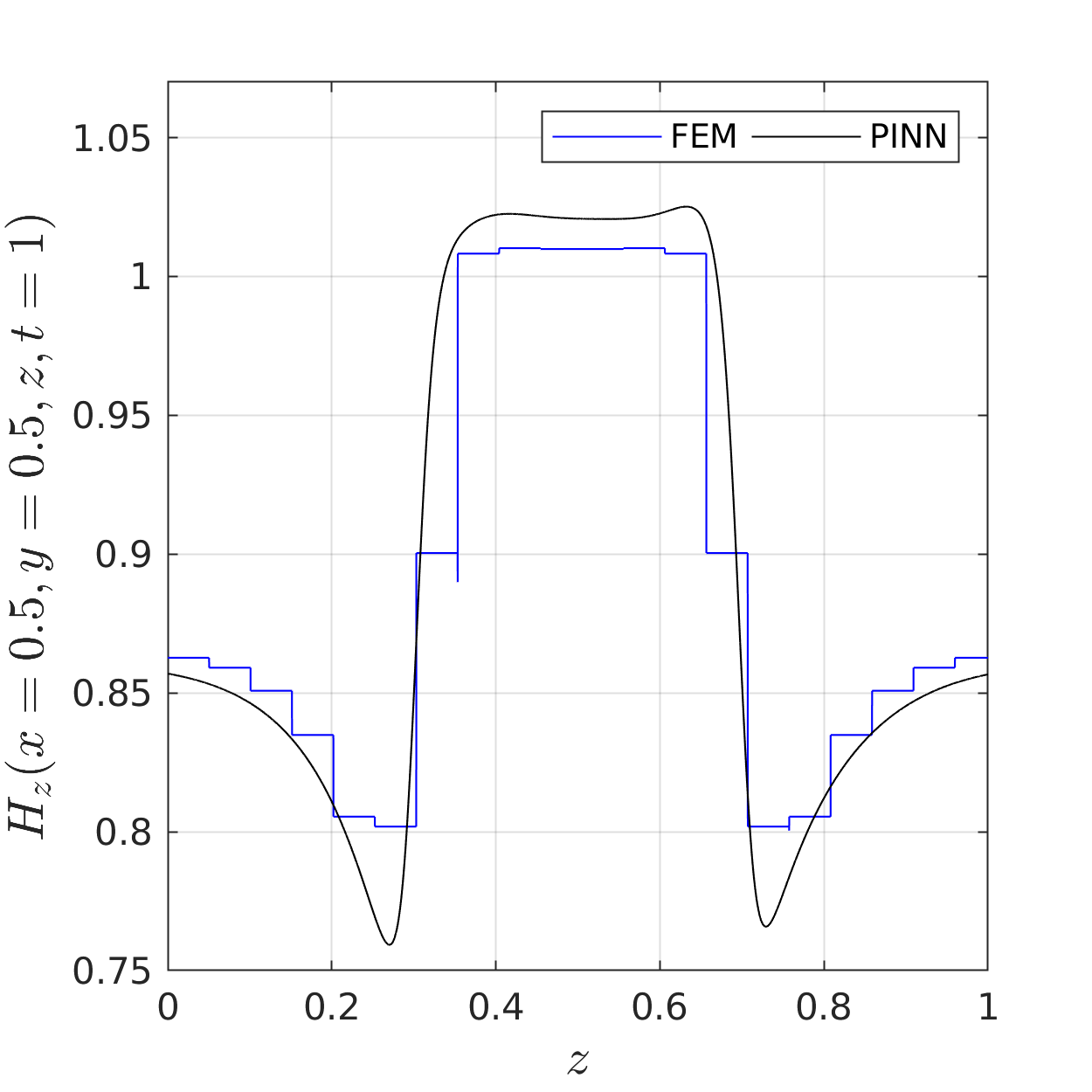}
    \caption{}
\end{subfigure}
    \caption{$z$-component of the magnetic field $\boldsymbol{H}_z$ for $x=y=0.5$, using the proposed PINN and the FEM for the transient electromagnetic, parametric problem with an interface: (a) $\mu_{\rm out}=0.5$, $t=0.5$; (b) $\mu_{\rm out}=0.5$, $t=1$; (c) $\mu_{\rm out}=1.5$, $t=0.5$; (d) $\mu_{\rm out}=1.5$, $t=1$. }
    \label{fig:tr_mu1}

  \end{figure}

Finally, we apply this numerical method to the transient, parametric electromagnetic problem with a discontinuous media. A sphere of radius $r=0.2$ and permeability $\mu =1$ nondimensional units is located inside a unit cube with permeability $\mu_{\rm out}$. The resistivity is $\rho =1$ on the whole domain, and the boundary conditions are $\boldsymbol{H}=[0,0,t]$ and $\boldsymbol{E}=[0,0,0]$. The time is $t \in [0,1]$ and the parameter $\mu_{\rm out} \in [0.5,1.5]$. We train the PINN using 4000 Adam iterations and 4000 L-BFGS iterations. We compare the results obtained using the PINN with the results obtained using the FEM in figure~\ref{fig:tr_mu1}.
The results obtained using the PINN are smoother than those obtained using the FEM. This is due to the inclusion of the divergence-free constraint in the training, as opposed to the FEM method, where the divergence-free condition is considered inherited from the initial solution. Breaking the divergence-free property is known to be the main reason for spurious solutions in computational electromagnetics~\cite{brackbill1980effect,toth2000b}.

\section{Conclusion}
In this work, we tackled three-dimensional, transient, and parametric electromagnetic problems with a material interface. We replaced the discontinuity at the interface with a smoothed Heaviside function that we built using a level-set function.
Among all the existing formulations of Maxwell's equations, we showed, using the NTK, that the first-order formulations are better suited for discontinuous problems, or problems with sharp gradients. Doing so alleviates the need to include the interface jump condition, which would naturally be satisfied if Maxwell's equations were satisfied. We proposed a neural network architecture that strongly imposes boundary and initial conditions, and incorporates sharp gradients at the interface. A numerical study was conducted to analyze the impact of the function $d(\boldsymbol{x})$ used for imposing the boundary condition. We discovered that to have optimal convergence, the derivative of $d(\boldsymbol{x})$ at the boundary must have the same order of magnitude as the derivatives of the exact solution, to ensure that the network's output is normalized. The proposed methodology was tested on numerous 3D problems, including parametric static problems and parametric time-dependent problems, and was found to give results comparable to the FEM method. The proposed architecture, however, does not scale well for problems with multiple interfaces, because the network size will grow proportionally with the number of interfaces. Another drawback of the proposed architecture is that it does not directly address the frequency bias of neural networks, so increasing the sharpness of the interface increases the training difficulties of the neural network.   

\clearpage
\bibliographystyle{abbrv} 
\bibliography{max}

\end{document}